\documentclass[
reprint,
superscriptaddress,
nofootinbib,
 amsmath,amssymb,
 aps,
]{revtex4-2}
\usepackage[pdftex]{graphicx}% Include figure files
\usepackage{epsfig} % convert eps to pdf
\usepackage{dcolumn}% Align table columns on decimal point
\usepackage{bm}% bold math
\usepackage[utf8]{inputenc}
\usepackage[T1]{fontenc}
\usepackage{mathptmx}
\usepackage{amsmath, amssymb}
\usepackage{hyperref}
\usepackage{color}
\usepackage{rotating}
\usepackage{natbib}
\usepackage{orcidlink}
\newcommand{\gluex}{GlueX }
\renewcommand{\Re}{\textrm{Re}}
\renewcommand{\Im}{\textrm{Im}}

\begin{document}

\title{Measurement of Spin-Density Matrix Elements in \texorpdfstring{$\boldsymbol{\rho}$(770)}{rho770} Production with a Linearly Polarized Photon Beam at \texorpdfstring{$\boldsymbol{E_\gamma = 8.2\,-\,8.8\,}\text{GeV}$}{Egamma=8.2--8.8GeV}}

\affiliation{Polytechnic Sciences and Mathematics, School of Applied Sciences and Arts, Arizona State University, Tempe, Arizona 85287, USA}
\affiliation{Department of Physics, National and Kapodistrian University of Athens, 15771 Athens, Greece}
\affiliation{Ruhr-Universit\"{a}t-Bochum, Institut f\"{u}r Experimentalphysik, D-44801 Bochum, Germany}
\affiliation{Helmholtz-Institut f\"{u}r Strahlen- und Kernphysik Universit\"{a}t Bonn, D-53115 Bonn, Germany}
\affiliation{Department of Physics, Carnegie Mellon University, Pittsburgh, Pennsylvania 15213, USA}
\affiliation{Department of Physics, The Catholic University of America, Washington, D.C. 20064, USA}
\affiliation{Department of Physics, University of Connecticut, Storrs, Connecticut 06269, USA}
\affiliation{Department of Physics, Duke University, Durham, North Carolina 27708, USA}
\affiliation{Department of Physics, Florida International University, Miami, Florida 33199, USA}
\affiliation{Department of Physics, Florida State University, Tallahassee, Florida 32306, USA}
\affiliation{Department of Physics, The George Washington University, Washington, D.C. 20052, USA}
\affiliation{School of Physics and Astronomy, University of Glasgow, Glasgow G12 8QQ, United Kingdom}
\affiliation{GSI Helmholtzzentrum f\"{u}r Schwerionenforschung GmbH, D-64291 Darmstadt, Germany}
\affiliation{Institute of High Energy Physics, Beijing 100049, People's Republic of China}
\affiliation{Department of Physics, Indiana University, Bloomington, Indiana 47405, USA}
\affiliation{National Research Centre Kurchatov Institute, Moscow 123182, Russia}
\affiliation{Department of Physics, Lamar University, Beaumont, Texas 77710, USA}
\affiliation{Department of Physics, University of Massachusetts, Amherst, Massachusetts 01003, USA}
\affiliation{Department of Physics, Massachusetts Institute of Technology, Cambridge, Massachusetts 02139, USA}
\affiliation{National Research Nuclear University Moscow Engineering Physics Institute, Moscow 115409, Russia}
\affiliation{Department of Physics, Mount Allison University, Sackville, New Brunswick E4L 1E6, Canada}
\affiliation{Department of Physics, Norfolk State University, Norfolk, Virginia 23504, USA}
\affiliation{Department of Physics, North Carolina A\&T State University, Greensboro, North Carolina 27411, USA}
\affiliation{Department of Physics and Physical Oceanography, University of North Carolina at Wilmington, Wilmington, North Carolina 28403, USA}
\affiliation{Department of Physics, Old Dominion University, Norfolk, Virginia 23529, USA}
\affiliation{Department of Physics, University of Regina, Regina, Saskatchewan S4S 0A2, Canada}
\affiliation{Departamento de Física, Universidad T\'ecnica Federico Santa Mar\'ia, Casilla 110-V Valpara\'iso, Chile}
\affiliation{Thomas Jefferson National Accelerator Facility, Newport News, Virginia 23606, USA}
\affiliation{Laboratory of Particle Physics, Tomsk Polytechnic University, 634050 Tomsk, Russia}
\affiliation{Department of Physics, Tomsk State University, 634050 Tomsk, Russia}
\affiliation{Department of Physics and Astronomy, Union College, Schenectady, New York 12308, USA}
\affiliation{Department of Physics, Washington \& Jefferson College, Washington, Pennsylvania 15301, USA}
\affiliation{Department of Physics, William \& Mary, Williamsburg, Virginia 23185, USA}
\affiliation{School of Physics and Technology, Wuhan University, Wuhan, Hubei 430072, People's Republic of China}
\affiliation{A. I. Alikhanyan National Science Laboratory (Yerevan Physics Institute), 0036 Yerevan, Armenia}
\author{S.~Adhikari} \affiliation{Department of Physics, Old Dominion University, Norfolk, Virginia 23529, USA}
\author{F.~Afzal\orcidlink{0000-0001-8063-6719 }} \affiliation{Helmholtz-Institut f\"{u}r Strahlen- und Kernphysik Universit\"{a}t Bonn, D-53115 Bonn, Germany}
\author{C.~S.~Akondi\orcidlink{0000-0001-6303-5217}} \affiliation{Department of Physics, Florida State University, Tallahassee, Florida 32306, USA}
\author{M.~Albrecht\orcidlink{0000-0001-6180-4297}} \affiliation{Thomas Jefferson National Accelerator Facility, Newport News, Virginia 23606, USA}
\author{M.~Amaryan\orcidlink{0000-0002-5648-0256}} \affiliation{Department of Physics, Old Dominion University, Norfolk, Virginia 23529, USA}
\author{V.~Arroyave} \affiliation{Department of Physics, Florida International University, Miami, Florida 33199, USA}
\author{A.~Asaturyan\orcidlink{0000-0002-8105-913X}} \affiliation{Department of Physics and Physical Oceanography, University of North Carolina at Wilmington, Wilmington, North Carolina 28403, USA}\affiliation{A. I. Alikhanyan National Science Laboratory (Yerevan Physics Institute), 0036 Yerevan, Armenia}
\author{A.~Austregesilo\orcidlink{0000-0002-9291-4429}} \email[Corresponding author: ]{aaustreg@jlab.org} \affiliation{Thomas Jefferson National Accelerator Facility, Newport News, Virginia 23606, USA}
\author{Z.~Baldwin\orcidlink{0000-0002-8534-0922}} \affiliation{Department of Physics, Carnegie Mellon University, Pittsburgh, Pennsylvania 15213, USA}
\author{F.~Barbosa} \affiliation{Thomas Jefferson National Accelerator Facility, Newport News, Virginia 23606, USA}
% ...
\collaboration{The \textsc{GlueX} Collaboration}

\date{\today} 
\begin{abstract}
The \gluex experiment at Jefferson Lab studies photoproduction of mesons using linearly polarized $8.5\,\text{GeV}$ photons impinging on a hydrogen target which is contained within a detector with near-complete coverage for charged and neutral particles. We present measurements of spin-density matrix elements for the photoproduction of the vector meson $\rho$(770). The statistical precision achieved exceeds that of previous experiments for polarized photoproduction in this energy range by orders of magnitude. We confirm a high degree of $s$-channel helicity conservation at small squared four-momentum transfer $t$ and are able to extract the $t$-dependence of natural and unnatural-parity exchange contributions to the production process in detail. We confirm the dominance of natural-parity exchange over the full $t$ range. We also find that helicity amplitudes in which the helicity of the incident photon and the photoproduced $\rho(770)$ differ by two units are negligible for $-t<0.5\,\text{GeV}^{2}/c^{2}$.
\end{abstract}
\maketitle

\section{Introduction}
The photoproduction of $\rho$(770) mesons off the proton is one of the photoproduction processes in which the spin state of the incident photon is conserved in the produced vector meson. The reaction can be described by the vector-meson-dominance model~\cite{Sakurai:1960ju} where the incident photon fluctuates into a vector meson (e.g. $\rho(770)$) which then interacts with the target nucleon. At beam energies well above $10\,\text{GeV}$, the process is expected to proceed through diffractive scattering with $s$-channel helicity conservation~\cite{Gilman:1970vi,Harari:1970fw,Bialas:1971xk} (SCHC). In order to describe this process, both the differential cross section for $\rho(770)$ photoproduction and the spin-density matrix elements (SDMEs) need to be measured. While the differential cross sections were extensively measured elsewhere~\cite{Schopper:1988hwx}, the SDMEs quantify the transfer of the photon spin state to that of the vector meson and most can only be accessed using a polarized photon beam. A detailed description of the SDMEs, and their connection to photoproduction, can be found in Ref.~\cite{Schilling:1969um}. More recently, Tabakin and colleagues have revisited the topic of vector-meson SDMEs in several different frameworks~\cite{Pichowsky:1994gh,Kloet:1998js,Kloet:1999wc}. With a beam of linearly polarized photons, nine real elements of the complex-valued spin-density matrix can be measured, and, in the case of SCHC, all but two of these should be zero when measured in the helicity system~(see~Sec.~\ref{sec:method}).

The first measurements of SDMEs in the photoproduction of $\rho(770)$ mesons with linearly polarized photons in the $1.4$ to $3.3~\text{GeV}$ energy range came from DESY~\cite{Criegee:1968fxl}. Their measurements of the beam asymmetry suggested nearly pure diffractive photoproduction over the entire energy range. A later measurement from Cornell using $3.5\,\text{GeV}$ linearly polarized photons also reported on the beam asymmetry, but saw some deviation from diffractive behavior~\cite{Diambrini-Palazzi:1970jgv}. Several measurements from SLAC with linearly polarized photons of energy $2.8$ and $4.7\,\text{GeV}$~\cite{Ballam:1970qn,Ballam:1971yd} and later including $9.3\,\text{GeV}$ photons~\cite{Ballam:1972eq} reported detailed SDMEs as well as agreement with SCHC and dominance of natural-parity exchange (NPE) in the production process (see~Appendix~\ref{sec:appschc} for a discussion of SCHC and NPE). Subsequent experiments at CERN with unpolarized $20$ to $70\,\text{GeV}$ photons measured the three unpolarized SDMEs~\cite{Bonn-CERN-EcolePoly-Glasgow-Lancaster-Manchester-Orsay-Paris-Rutherford-Sheffield:1982qiv}. Finally, measurements with the Hybrid Bubble Chamber facility at SLAC measured the $\rho(770)$ SDMEs with $20\,\text{GeV}$ linearly polarized photons~\cite{SLACHybridFacilityPhoton:1984hkg}. While of limited statistical precision, all previous measurements are consistent with a dominance of natural-parity exchange and show that SCHC is valid at least over a limited range in momentum transfer $t$ (see~Appendix~\ref{sec:appschc}).

The Joint Physics Analysis Center (JPAC) has recently developed a model based on Regge theory amplitudes to describe the photoproduction of light vector mesons~\cite{Mathieu:2018xyc}. JPAC fitted this model to the SLAC results and other cross section measurements, and produced theoretical predictions for the spin-density matrix elements at $8.5\,\text{GeV}$. According to the prediction, the dominant contributions to the photoproduction of the $\rho$(770) meson at this beam energy stem from Pomeron and $f_2(1270)$ exchanges. The analytical form of this model uses an expansion in $\sqrt{-t}/(m_0c)$ where $m_0$ is the mass of the vector meson. Since it only takes into account the leading terms of this expansion, we limit the comparison with our data to $-t < m^2_0c^2 \approx 0.5\,\text{GeV}^{2}/c^{2}$ even though our results cover a larger range in $t$.

Sections~\ref{sec:experiment}, \ref{sec:data} and~\ref{sec:acceptance} describe the experimental setup and data collection, the selection of $\rho(770)$ production events from the data and the determination of the detector's acceptance. Section~\ref{sec:analysis} sets out the details of the analysis: it shows how the spin-density matrix elements are obtained from the angular distribution of the $\rho(770)$'s decay products, and describes the fit method and the measurements' uncertainties.  Section~\ref{sec:results} presents and discusses the results.  In Appendix~\ref{sec:appschc} we discuss s-channel helicity conservation and its implications for spin-density matrix elements of vector-meson states produced by natural-parity exchange. The measurements presented in this article supersede preliminary \gluex results~\cite{Austregesilo:2019tld}.

\section{The \gluex Experiment}\label{sec:experiment}
 The \gluex experiment~\cite{GlueX:2020idb} at the Thomas Jefferson National Accelerator Facility is part of a global effort to study the spectrum of hadrons. A primary electron beam with an energy of up to $12\,\text{GeV}$ is used to produce a secondary photon beam which impinges on a liquid-hydrogen target. The scattered electrons tag the energy of the beam photons. A high beam intensity provides a sufficiently large reaction rate to study rare processes. The \gluex detector has been specifically designed to map the light-quark meson spectrum up to masses of approximately $3\,\text{GeV}/c^2$ with full acceptance for all decay modes. A $2$~T superconducting solenoid houses the target, a start counter~\cite{Pooser:2019rhu}, central~\cite{Jarvis:2019mgr} and forward drift chambers~\cite{Pentchev2017281}, and a barrel calorimeter~\cite{Beattie:2018xsk}. A forward calorimeter completes the forward photon acceptance and a time-of-flight counter provides particle identification capability.

The key feature of \gluex is its capability to use a polarized photon beam. Linear polarization of the photons is achieved by coherent bremsstrahlung of the primary electron beam on a thin diamond radiator. With a collimator reducing the contribution from the incoherent bremsstrahlung spectrum, a degree of linear polarization of up to $35\%$ is achieved in the coherent peak at $8.8\,\text{GeV}$. In order to cancel apparatus effects, data are collected with the polarization plane in four different orientations, rotated about the beam direction in steps of $45^{\circ}$. The degree of polarization is measured using the triplet production effect~\cite{Dugger:2017zoq}. As the primary electron beam helicity is flipped pseudo-randomly multiple times per second, the circularly polarized component of the photon beam is averaged out.

The photon beam polarization imposes constraints on the properties of the production process. It may be used as a filter to enhance particular resonances or as an additional input to multidimensional amplitude analyses. To this end, the photoproduction mechanism must be understood in great detail. Only very limited data from previous experiments are available at these energies. \gluex has already measured beam-asymmetry observables for the production of several pseudoscalar mesons: $\gamma p\rightarrow \pi^0 p$~\cite{GlueX:2017zoo}, $\gamma p\rightarrow \eta p$ and $\gamma p\rightarrow \eta^{\prime}(958) p$~\cite{GlueX:2019adl}, $\gamma p \rightarrow  K^{+} \Sigma^{0}$~\cite{GlueX:2020qat}, and $\gamma p \rightarrow \pi^- \Delta^{++}(1232)$~\cite{GlueX:2020fam}. In addition to the beam-asymmetry measurements, we have also reported SDMEs for the photoproduction of the $\Lambda(1520)$~\cite{GlueX:2021pcl}. As an extension of this program, the following analysis studies the production process for the $\rho$(770) vector meson.

The first phase of the \gluex experiment, consisting of three run periods, recorded a total integrated luminosity in the coherent peak of about $125\,\text{pb}^{-1}$. Only the data from the first of those run periods (about $17\%$ of the full data set) are used to produce the results discussed here.

\section{Description of Data Set}\label{sec:data}

\begin{figure}[ht]\centering
        \includegraphics[width=0.5\textwidth]{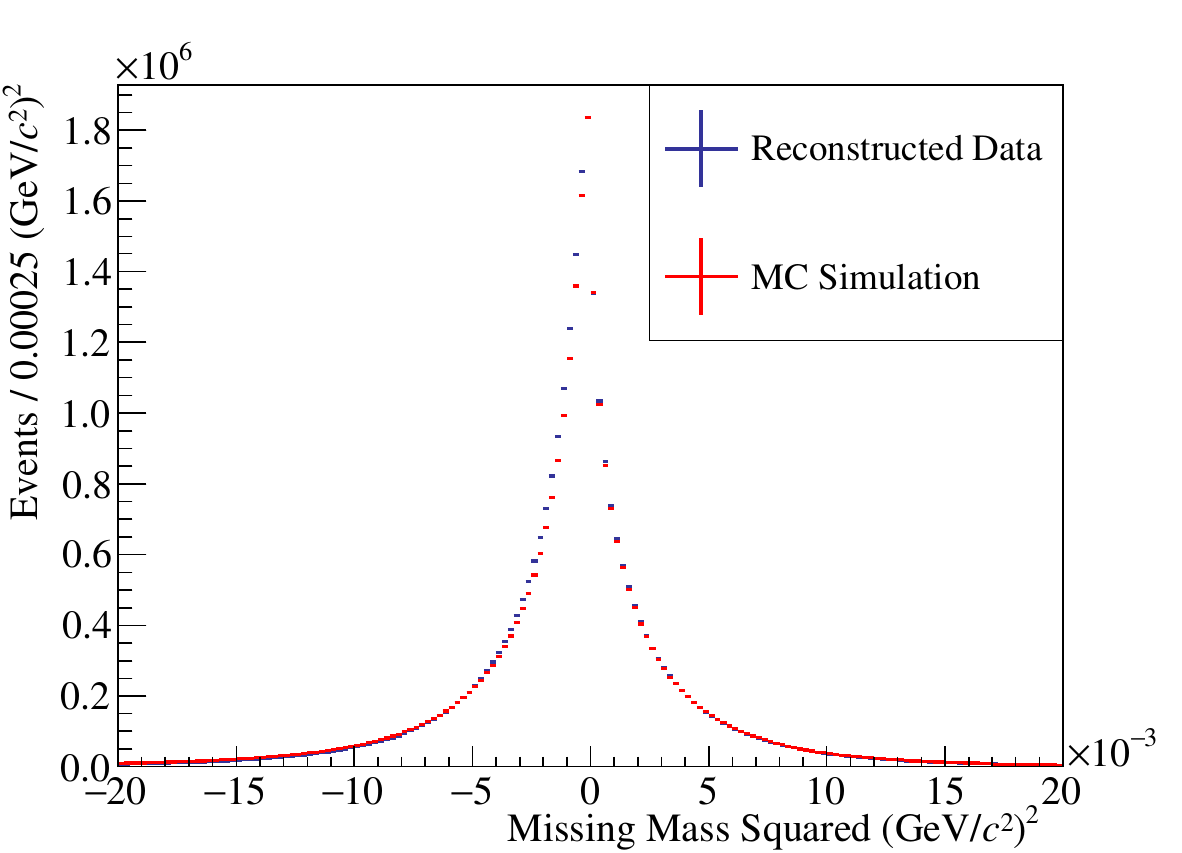}
		\caption{The squared missing mass distribution from the reaction $\gamma p \rightarrow \pi^{+}\pi^{-} X_\text{miss} p$ calculated using the values of momentum and energy of the final-state particles before they are constrained by the kinematic fit.}
       \label{fig:mm_mc}
\end{figure}

We study the reaction $\gamma p \rightarrow \rho(770) p$, where the $\rho$(770) meson decays predominantly into the $\pi^+\pi^-$ final state~\cite{ParticleDataGroup:2022pth}. We select exclusive events by completely reconstructing the final state $\pi^+\pi^-p$ with all particle trajectories originating from the same vertex. A seven-constraint kinematic fit is performed on each event, which enforces energy and momentum conservation for the reaction $\gamma p\rightarrow \pi^{+}\pi^{-}p$ as well as a common vertex for all particles. We accept only events where the kinematic fit converges with $\chi^2/\text{ndf}<5.0$, which removes backgrounds originating from misidentified charged tracks and non-exclusive events. The final event selection is applied for all figures in this section. Figure~\ref{fig:mm_mc} shows the squared missing mass from the assumed reaction $\gamma p \rightarrow \pi^{+}\pi^{-} X_{\text{miss}}p$ calculated using the values of momentum and energy of the final-state particles before they are constrained by the kinematic fit. The observed peak very close to zero implies that there are no massive missing particles.

The $\pi^+\pi^-p$ final state measured by the \gluex detector is matched to the initial state photon via its energy and timing. Due to the large incoming photon flux and limited resolution, accidental coincidences can fulfill the matching requirement and contaminate the event sample. The primary electron beam is produced with a 250\,MHz time structure, which translates into photon beam bunches that are $4\,\text{ns}$ apart. We estimate the accidental background by intentionally selecting events from neighboring beam bunches. In this analysis, we select four beam bunches on each side of the prompt signal peak as side band regions and weight those events by $-\tfrac{1}{8}$ to achieve similar statistical precision for signal and background. About 20\% of the events are statistically subtracted from the signal sample with this method.

\begin{figure}[ht] \centering
		\includegraphics[width=0.5\textwidth]{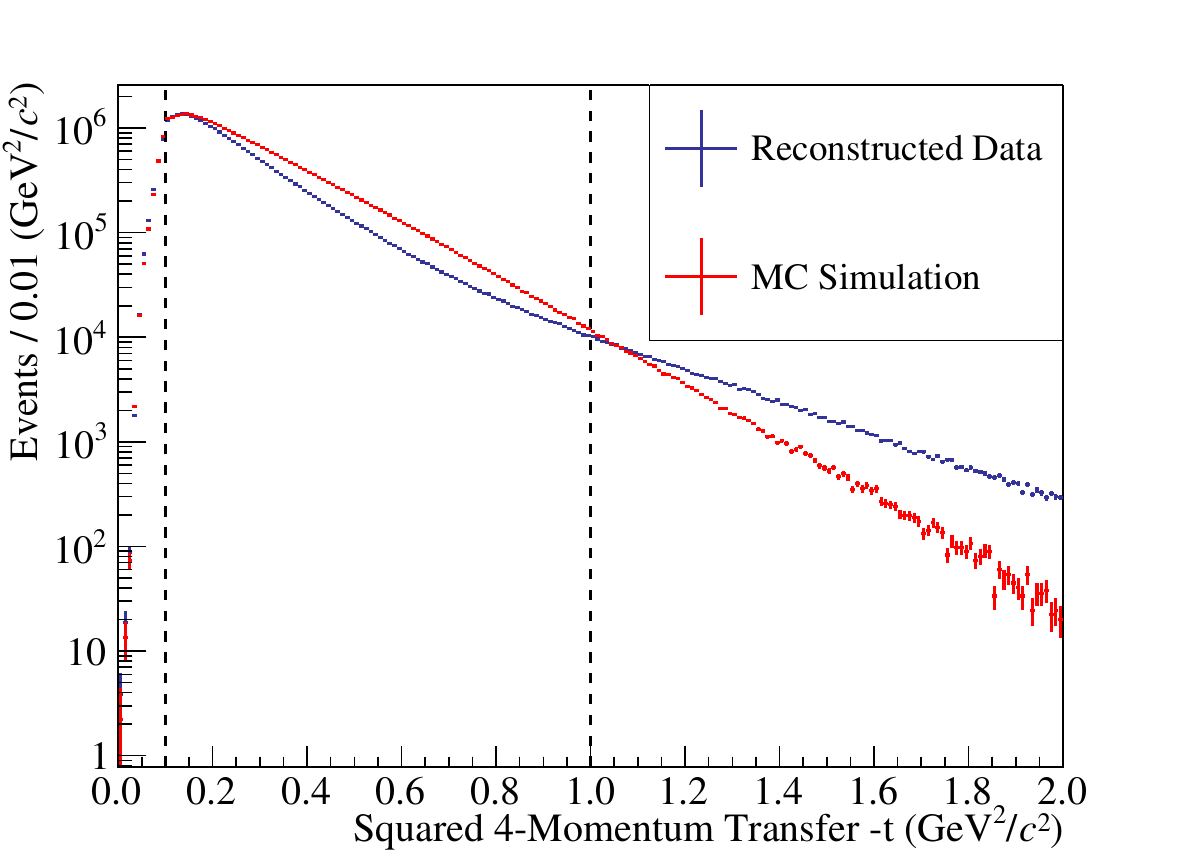}
		\caption{The distribution of the squared four-momentum transfer $-t$. The dashed vertical lines indicate the range analyzed.}
		\label{fig:t_mc}
    \end{figure}

 Due to the requirement for a successfully reconstructed proton track, the distribution of the squared four-momentum transfer $t$ shows a depletion at zero (see~Fig.~\ref{fig:t_mc}). Since the acceptance is very low in this region, we discard all events with $-t$ below $0.1\,\text{GeV}^{2}/c^{2}$. Above $-t=1\,\text{GeV}^2/c^2$, the slope of the distribution has changed visibly, which indicates a deviation from a simple $t$-channel process. To avoid effects from potential target excitation, we limit the analysis to the region below this value of $-t$.

\begin{figure}[ht] \centering
		\includegraphics[width=0.5\textwidth]{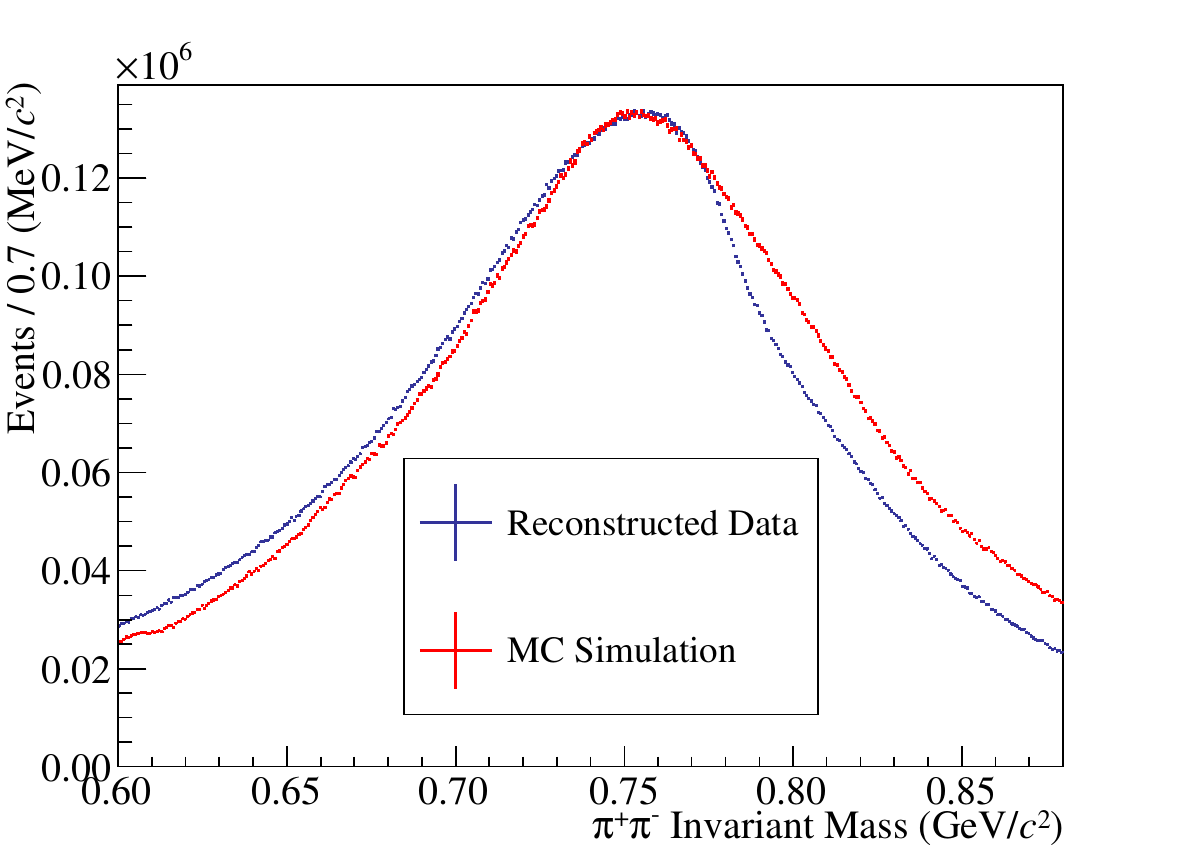}
	    \caption{The invariant mass distribution of the produced $\pi^+\pi^-$ system. The small differences between data and simulation are due to non-resonant background under the $\rho(770)$ and interference with the decay $\omega(782)\rightarrow\pi^+\pi^-$; neither of these are present in the simulation. For further analysis, the simulated events are re-weighted in order to match the mass distribution of the measured data exactly.}
	    \label{fig:m2pi_mc}
\end{figure}

 We separate the $\rho$(770) meson signal from the continuous $\pi^+\pi^-$ spectrum by selecting the invariant mass of the di-pion system to be between $0.60$ and $0.88\,\text{GeV}/c^{2}$. This selection suppresses non-$\pi^+\pi^-$ background to an almost negligible amount, but is not able to distinguish the $\rho$(770) resonance from contributions from non-resonant $\pi^+\pi^-$ production. A phenomenological fit to the invariant mass distribution is used to estimate the fraction of non-resonant production to be less than 1\% of the total. As this fraction is strongly $t$-dependent and approaches 10\% above $t=1\,\mathrm{GeV}^2/c^2$, this analysis is limited to $t<1\,\mathrm{GeV}^2/c^2$. It is well known that the interference between the $\rho(770)$ resonance and the underlying non-resonant background can shift the apparent mass of the vector meson~\cite{Soding:1965nh}. We observe the $\rho(770)$ peak in the $\pi^+\pi^-$ mass distribution (see~Fig.~\ref{fig:m2pi_mc}) about $18\,\text{MeV}/c^{2}$ below the PDG average for the mass of the photoproduced neutral $\rho(770)$, which is $769.2\pm0.9\,\text{MeV}/c^2$~\cite{ParticleDataGroup:2022pth}.

A simulation of possible background channels indicates that the contribution of final states other than exclusive $\pi^+\pi^-$ production is negligible, at less than 1 in 1000. This study also shows that the decay $\omega(782)\rightarrow\pi^+\pi^-$ constitutes an irreducible background component. As the decay is suppressed by $G$-parity, it only amounts to approximately 0.4\% of the data sample. This agrees with the estimation from known cross sections and branching fractions~\cite{ParticleDataGroup:2022pth} and has no measurable impact on the presented results.

 In total, we obtain data samples with nearly $9\times10^6$ $\rho(770)$ candidate events for each of the four orientations of the beam-photon polarization. We extract the spin-density matrix elements in $18$ bins of $-t$ between $0.1$ and $1.0\,\text{GeV}^{2}/c^{2}$. We use a logarithmic function to determine the bin boundaries so that the number of events in each bin is approximately equal.


\section{Simulation of Detector Acceptance}
\label{sec:acceptance}

 To extract the spin-density matrix elements of the $\rho(770)$ from the measured angular distribution of its decay products, we must correct for acceptance effects. The acceptance of the \gluex detector has been simulated based on a Geant4~\cite{ALLISON2016186} detector model, with a subsequent smearing step to reproduce the resolution effects of the individual detector subsystems.  Detailed comparisons between the simulation and measurements have been reported elsewhere~\cite{GlueX:2020idb}.
 
 We simulate a signal sample that reproduces the production kinematics of the measured process, but has an isotropic distribution in the decay angles. To describe the process $\gamma p\rightarrow \pi^{+}\pi^{-}p$,  we assume an exponential distribution of the squared four-momentum transfer, i.e. proportional to $e^{bt}$ with the slope parameter $b=6\,(\text{GeV}/c)^{-2}$. This simplified model does not reproduce the experimentally observed $t$-distribution exactly~(see~Fig.~\ref{fig:t_mc}), but serves as a good approximation when binning finely in $t$. We model the shape of the $\pi^+\pi^-$ invariant mass distribution in the range between $0.60$ and $0.88\,\textrm{GeV}/c^{2}$ using a relativistic $P$-wave Breit-Wigner~\cite{Breit:1936zzb} function with an orbital angular momentum barrier factor $F$ that is parameterized according to Ref.~\cite{hip72}: 
\begin{eqnarray}
BW(m) & = & \frac{\sqrt{m_0\Gamma_0}}{m^{2}-m^{2}_{0} - i m_0 \Gamma(m,L)} \\
\text{with}~~~\Gamma(m,L) & = & \Gamma_0 \frac{q}{m} \frac{m_0}{q_0} \left [ \frac{F(q, L)}{F(q_0, L)} \right ]^2.
\end{eqnarray}
Here, $q$ signifies the breakup momentum of the pions and $q_{0}$ is the breakup momentum at the nominal resonance mass $m_{0}$. The reconstructed mass distribution from the Monte Carlo simulation approximates the experimentally measured one with the parameters $m_{0} = 757\,\textrm{MeV}/c^{2}$ and $\Gamma_0=146\,\textrm{MeV}/c^{2}$~(see~Fig.~\ref{fig:m2pi_mc}). In a second step, the simulated sample is re-weighted in order to match the mass distribution of the measured data exactly.

\section{Analysis Method}
\label{sec:analysis}

We use an unbinned extended-maximum-likelihood fit to extract the spin-density matrix elements from the measured distribution. This method is widely used in amplitude analysis and has the advantage that neither the data nor the acceptance corrections have to be divided into regions of angular phase space.

\subsection{Spin-Density Matrix Elements}
\label{sec:method}
We characterize the photoproduction of vector mesons by an amplitude $T$, which connects the spin-density matrix $\rho(\gamma)$ for the initial photon beam to the spin-density matrix $\rho(V)$ of the produced vector meson. Following Schilling et al.~\cite{Schilling:1969um}, we write
\begin{eqnarray}
\rho(V) & = & T\, \rho(\gamma)\, T^{*} \, .
\label{eq:rhoTgam}
\end{eqnarray}
We can incorporate the photon polarization into the description of the vector-meson density matrix. The spin-density matrix for the photon can be written as
\begin{eqnarray}
    \rho(\gamma) & = & \frac{1}{2}\,I + \frac{1}{2}\,\mathbf{P}_{\gamma}\,\cdot\, \mathbf{\sigma}\, ,
\end{eqnarray}
where $I$ is the identity matrix, $\mathbf{\sigma}$ are the Pauli matrices and the vector $\mathbf{P}_{\gamma}$ is given as  
\begin{eqnarray}
\label{eq:photon_pol_lin}
\mathbf{P}_{\gamma} & = & P_{\gamma} \left ( -\cos 2\Phi, -\sin 2\Phi, 0 \right )\, ,
\end{eqnarray}
where $P_{\gamma}$ is the degree of linear polarization (between $0$ and $1$) and $\Phi$ is the angle between the polarization vector of the photon and the production plane of the vector meson.
In the case of circularly polarized photons,
\begin{eqnarray}
\label{eq:photon_pol_circ}
\mathbf{P}_{\gamma} & = & P_{\gamma} \left ( 0, 0, \lambda_\gamma \right )\, ,
\end{eqnarray}
where $P_{\gamma}$ is again the degree of polarization, and $\lambda_\gamma = \pm 1$ corresponds to the helicity of the photon. If we now consider the three components of the photon polarization (components $1$ and $2$ for linear polarization and component $3$ for circular  polarization), we can write the vector-meson density matrix as the sum 
\begin{eqnarray}
\label{eq:rho-alpha}
\rho(V) & = & \rho^{0} \, + \, \sum_{\alpha=1}^{3}\, P^{\alpha}_{\gamma}\, \rho^{\alpha} \, ,
\end{eqnarray}
where the $\rho^{\alpha}$ parameterize the dependence of the total density matrix on the photon polarization. Since we use a linearly polarized photon beam, we will ignore the contribution from circularly polarized photons in the remaining text by setting $\rho^{3} = 0$.

\begin{figure*}[ht!]\centering
        \includegraphics[trim={145 400 190 125}, clip, width=.45\textwidth]{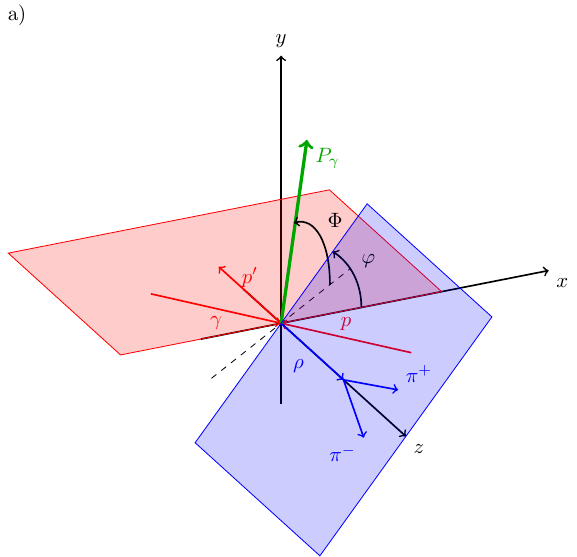}       
        \includegraphics[trim={145 400 190 125}, clip, width=.45\textwidth]{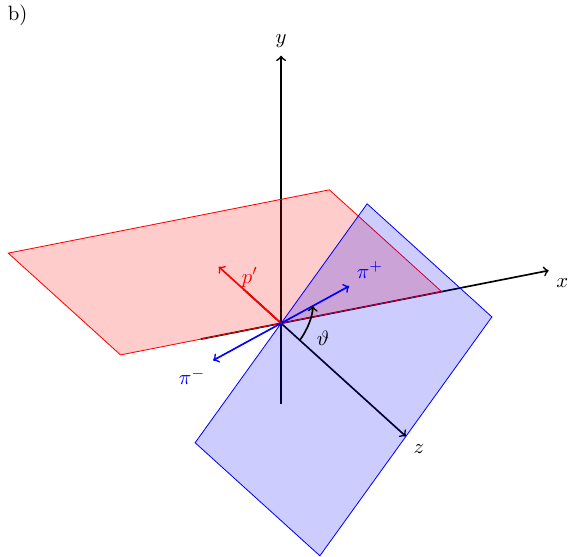}
        \caption[]{\label{fig:coordinates}Definition of the angles used to describe vector-meson photoproduction. The hadronic production plane and the $\rho$(770) decay plane are shown in red and blue, respectively. The photon polarization vector $P_\gamma$ is indicated in green. Diagram a) is in the center-of-mass frame of the reaction with the $z$ axis along the direction of the $\rho(770)$ meson; b) is boosted into the rest frame of the $\rho(770)$ meson, i.e. the helicity system.}
\end{figure*}

The spin-density matrix elements $\rho_{ij}^k$ in Eq.~\eqref{eq:rho-alpha} describe the angular dependence of the cross section. The number density $n$ of produced events in the experiment is proportional to the normalized angular distribution $W$, i.e.:
 \begin{equation}
 \label{eq:number_density}
     n(\vartheta, \varphi, \Phi) \propto W(\vartheta, \varphi, \Phi) \, .
 \end{equation}
 Here, $W$ is a function of the two decay angles $\vartheta$ and $\varphi$, defined in the helicity system of the vector meson (see~Fig.~\ref{fig:coordinates}), and $\Phi$, the direction of the photon polarization with respect to the hadronic production plane as determined in the center-of-mass frame of the reaction. Together with the independently measured degree of polarization $P_{\gamma}$, the angular distribution for vector-meson production with a linearly polarized photon beam can be written as follows:

\begin{eqnarray}
 W(\cos\vartheta, \varphi, \Phi) =&& W^{0}(\cos\vartheta, \varphi) - P_\gamma \cos(2\Phi) W^{1}(\cos\vartheta,\varphi) \nonumber \\ & - &  P_\gamma\sin(2\Phi)W^{2}(\cos\vartheta, \varphi) \, .
 \label{eq:sdme}
 \end{eqnarray}
 For the case of the vector meson decaying to two spinless particles, such as $\rho(770)\rightarrow \pi^+\pi^-$, the decay distributions $W^i(\cos\vartheta, \varphi)$ in Eq.~\eqref{eq:sdme} are given by
 \begin{eqnarray}
        W^0(\cos\vartheta, \varphi) &=& \frac{3}{4\pi} \left ( \frac{1}{2}(1-\rho^0_{00}) + \frac{1}{2}(3\rho^0_{00}-1)\cos^2\vartheta \right. \\
        & - & \left. \sqrt{2}\mathrm{Re}\rho^0_{10}\sin2\vartheta\cos\varphi-\rho^0_{1-1}\sin^2\vartheta\cos2\varphi \right ) \nonumber \\
        W^{1}(\cos\vartheta, \varphi) &=& \frac{3}{4\pi} \left ( \rho^1_{11}\sin^2\vartheta + \rho^1_{00}\cos^2\vartheta \right.  \\ & - &  \left .\sqrt{2}\mathrm{Re}\rho^1_{10}\sin2\vartheta\cos\varphi - \rho^1_{1-1}\sin^2\vartheta\cos2\varphi \right )\nonumber \\
        W^2(\cos\vartheta, \varphi) & = & \frac{3}{4\pi} \left ( \sqrt{2}\mathrm{Im}\rho^2_{10}\sin2\vartheta\sin\varphi \right. \label{eq:sdme2} \\ & + & \left. \mathrm{Im}\rho^2_{1-1}\sin^2\vartheta\sin2\varphi \right ) \, . \nonumber
\end{eqnarray}

\subsection{Unbinned Extended Maximum Likelihood Fit}
\label{sec:fit_maths}

The agreement between the measured event distribution and the acceptance-weighted model given in Eqs.~\eqref{eq:number_density} to \eqref{eq:sdme2} is optimized by varying the spin-density matrix elements $\rho^i_{jk}$ and an external normalization factor $K$ as fit parameters. For this purpose, the extended likelihood function is maximized by a numerical algorithm. For the construction of this likelihood function, the probability for an event $i$ characterized by $\vartheta_i$, $\varphi_i$ and $\Phi_i$ to be observed by the experiment with acceptance $\eta(\vartheta, \varphi, \Phi)$ is defined by
\begin{equation}
  P_i = \frac {n(\vartheta_i, \varphi_i, \Phi_i)\eta(\vartheta_i, \varphi_i, \Phi_i)} {\int \mathrm{d}\cos\vartheta \mathrm{d}\varphi \mathrm{d}\Phi\, n(\vartheta, \varphi, \Phi) \eta(\vartheta, \varphi, \Phi)}~.
  \label{eq:prob}
\end{equation}
The total number of observed events $N$ in an experiment of fixed duration follows the Poisson distribution with an expectation value $\bar N$. The extended likelihood function
\begin{equation}
  \mathcal{L} = \frac {e^{-\bar N} \bar N^{N}} {N!} \prod_{i=1}^N P_i~
  \label{eq:poisson}
\end{equation}
takes this variation into account. The expectation value $\bar N$ is identical to the integral in the denominator of Eq.~\eqref{eq:prob}:
\begin{equation}
    \bar N  = \int \mathrm{d}\cos\vartheta \mathrm{d}\varphi \mathrm{d}\Phi\, n(\vartheta, \varphi, \Phi) \eta(\vartheta, \varphi, \Phi) ~.
    \label{eq:Nbar}
\end{equation}
Hence, the likelihood function simplifies to
\begin{equation}
  \mathcal{L} = \frac {e^{-\bar N}} {N!} \prod_{i=1}^N n(\vartheta_i, \varphi_i, \Phi_i)\eta(\vartheta_i, \varphi_i, \Phi_i)~.
\end{equation}

As large sums are computationally easier to handle than large products, we maximize the logarithm of the likelihood function
\begin{eqnarray}
  \ln \mathcal{L} & = & \sum_{i=1}^N \ln n(\vartheta_i, \varphi_i, \Phi_i) + \underbrace{\sum_{i=1}^N \ln \eta(\vartheta_i, \varphi_i, \Phi_i) - \ln N!}_{\mathrm{const}} \nonumber \\ & & - \int \mathrm{d}\cos\vartheta \mathrm{d}\varphi \mathrm{d}\Phi\, n(\vartheta, \varphi, \Phi) \eta(\vartheta, \varphi, \Phi)
  \label{eq:likelihood}
\end{eqnarray}
 in order to find the model parameters that match best the observed angular distribution $n(\vartheta, \varphi, \Phi)$.
The constant terms $\Sigma \ln \eta$ and $\ln N!$ do not depend on the fit parameters and can therefore be omitted from the fit. The recorded data sample only appears in the first sum, where events from neighboring beam bunches enter with negative weights to subtract background from accidental beam coincidences. The so-called normalization integral that contains the experimental acceptance is evaluated using the large phase-space Monte Carlo sample introduced in section~\ref{sec:acceptance}. This allows us to separate the normalization factor from the SDME fit parameters:
\begin{eqnarray}
  \int \mathrm{d}\cos\vartheta \mathrm{d}\varphi \mathrm{d}\Phi\, n(\vartheta, \varphi, \Phi) \eta(\vartheta, \varphi, \Phi) & = & \nonumber \\   K \underbrace{\int \mathrm{d}\cos\vartheta \mathrm{d}\varphi \mathrm{d}\Phi\, W(\vartheta, \varphi, \Phi) \eta(\vartheta, \varphi, \Phi)}_{\mathbb{I}} \, .& & 
\end{eqnarray}
The normalization integral $\mathbb{I}$ is approximated by summing over all generated phase-space events $N_\text{MC}^{\text{acc}}$ that pass the reconstruction and selection criteria after the detector simulation:
\begin{equation}
  \mathbb{I} \approx \frac {8\pi^2} {N_{\mathrm{MC}}} \sum_{j=1}^{N_{\mathrm{MC}}^{\mathrm{acc}}} W(\vartheta_j, \varphi_j, \Phi_j)\, ,
  \label{eq:normalization_integral}
\end{equation}
 where $N_\mathrm{MC}$ is the total number of generated Monte-Carlo events. The factor $8\pi^2$ is the integration volume.

The extended likelihood function is maximized by choosing the SDMEs as well as the normalization coefficient $K$ such that $n(\vartheta, \varphi, \Phi)$ matches the measured data best. This formalism has been implemented using the AmpTools software framework~\cite{amptools}. In contrast to conventional mass-independent amplitude analyses, the normalization integral depends on the fitted parameters, i.e. the SDMEs, and has to be recalculated at every iteration of the fit, with significant computational cost. For this reason, it was essential to use graphical processing units for the numerical evaluation of the large sums in Eqs.~\eqref{eq:likelihood} and \eqref{eq:normalization_integral}, which can contain up to $10^6$ summands in this analysis.

\subsection{Fit Evaluation}\label{sec:fit_evaluation}

For converged fits, we can evaluate the quality of the model with the expectation value $\bar N$ in Eq.~\eqref{eq:Nbar}. Using the numerical approximation of the normalization integral in Eq.~\eqref{eq:normalization_integral}:
\begin{equation}
    \bar N \approx  \frac{8\pi^2}{N_{\mathrm{MC}}} \sum_{j=1}^{N_{\text{MC}}^{\mathrm{acc}}} K W(\vartheta_j, \varphi_j, \Phi_j)~,
\end{equation}
we see that an individual MC event with the phase-space coordinates $(\vartheta_i, \varphi_i, \Phi_i)$ contributes with a weight:
\begin{equation}
  w_i = \frac{8\pi^2}{N_{\mathrm{MC}}} K W(\vartheta_i, \varphi_i, \Phi_i) \,
\end{equation}
to the data sample.
Events rejected by the reconstruction and kinematic selection have zero weight. The acceptance of the apparatus is therefore taken into account by construction. By applying these weights to the phase-space MC events, we obtain weighted MC samples that we can use to compare any kinematic distribution of the fitted model with the data. If the distributions of the angles that the model depends upon agree within statistical uncertainties, this would be a confirmation that the SDME model is sufficient to describe the data. The distributions in other kinematic variables can be used to assess how realistically the simulation reproduces detector effects.

\begin{figure*}[t]
\centerline{\includegraphics[width=.9\textwidth]{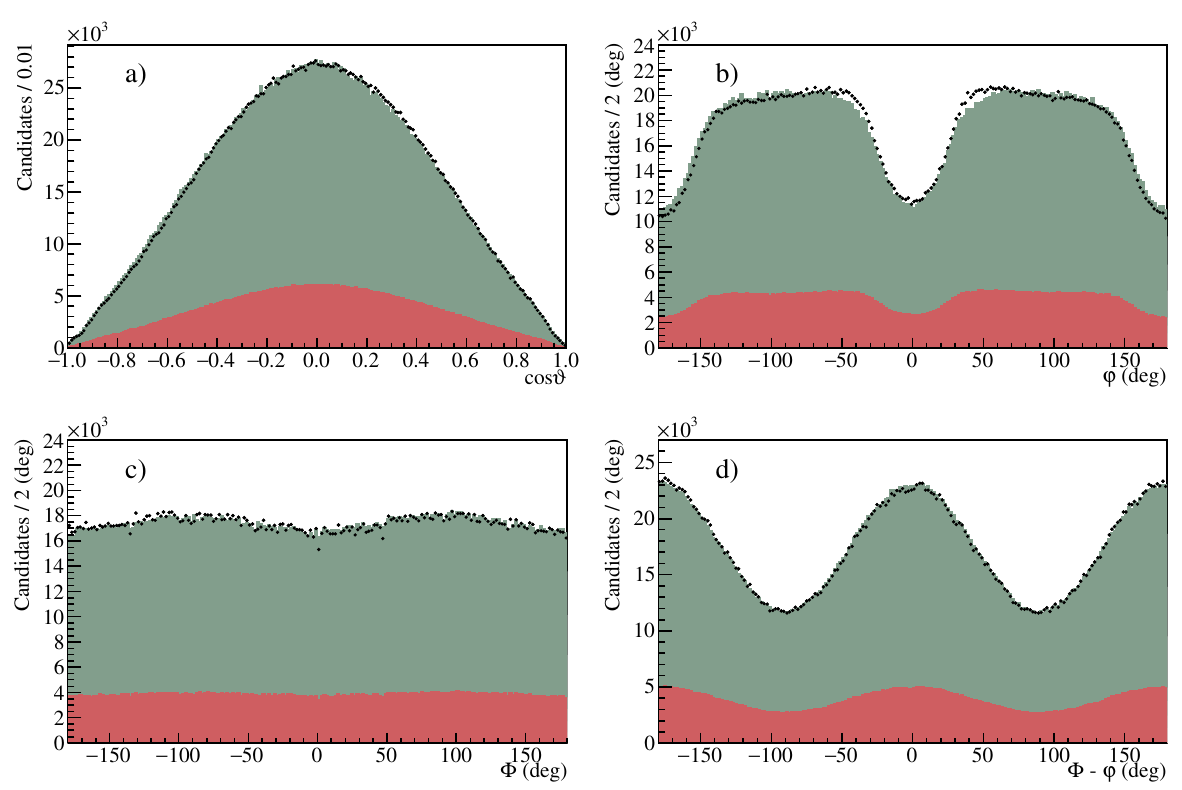}}
\caption{Evaluation of the fit by comparison of measured distributions (black) to phase-space simulation weighted with fit results (shaded green). The smaller contribution from the subtracted accidental background is  shown in red. Panel~a) shows the comparison for the cosine of the helicity angle $\vartheta$ and b) compares the distribution of the helicity angle $\varphi$. Panel~c) compares the azimuthal angle $\Phi$ of the polarization vector with respect to the production plane in the center-of-mass frame and d) shows the distribution of the difference between $\Phi$ and $\varphi$.  \label{fig:eval}} 
\end{figure*}

Figure~\ref{fig:eval} shows such a comparison for the combined fit of four orientations in one example bin at around $-t\approx0.2\,\text{GeV}^2/c^2$.  The distributions in the angles $\cos\vartheta$, $\varphi$, $\Phi$ and $\Phi-\varphi$ are very well reproduced.  A small asymmetry between $\varphi=0$ and $\varphi=\pm\pi$ indicates the possible interference with a $\pi\pi$ $S$-wave component, which is not included in the description of the system with vector-meson spin-density matrix elements.

\subsection{Discussion of Uncertainties}\label{sec:uncertainties}

We evaluate the statistical uncertainties with the Bootstrapping technique~\cite{efron1994introduction}. The analysis is repeated many times, each time using a different random sample of the same number of events selected from the original data, where some events are included more than once and others are omitted. We draw $200$ such samples and perform fits in the same way as for the real sample, keeping the starting values fixed at the nominal result. The distributions for the $9$ spin-density matrix elements from the $200$ fits can be well approximated by Gaussian functions, and their standard deviations serve as a measure of the statistical uncertainties.

A study of many possible sources for systematic uncertainties indicates that the only significant contributions arise from the beam polarization measurement and the selection of the signal sample. In particular, it is evident that the fitting procedure does not introduce any bias into the measured SDMEs and that there is no significant dependence of the SDMEs on the beam energy within the range studied.

The largest contribution to the systematic uncertainty originates from the external measurement of the beam-photon polarization.
The $1.5\%$ systematic uncertainty inherent in the design and the operation of the triplet polarimeter instrument~\cite{Dugger:2017zoq} is combined with the statistical uncertainty of the number of detected triplet events to give a total uncertainty of $2.1\%$. This overall normalization uncertainty is fully correlated for all bins in $t$. It is added in quadrature to the final uncertainties for the SDMEs $\rho^1_{ij}$ and $\rho^2_{ij}$, shown in Fig.~\ref{fig:rhosdme}, whose extraction is dependent on the polarization.

Small contributions to the systematic uncertainties are also caused by the selection of the signal sample, and they may have different magnitudes for each SDME and in each bin in $t$. To assess them, the requirements such as the convergence criterion of the kinematic fit or the suppression of possible background from excited baryons are varied such that the total event sample size does not change by more than $10\%$. The standard deviation for each type of variation is used as a measure of its systematic effect. If significant, the deviations are included in the quadratic sum, which we quote as the total systematic uncertainty for each data point individually. On average, the event selection adds about $2\cdot10^{-3}$ to the absolute value of the systematic uncertainty.% All other event selection criteria do not add significant systematic uncertainties to the results.

\section{Results}
\label{sec:results}

\subsection{\label{sec:sdme-final}Spin-Density Matrix Elements}

\begin{figure*}[pht!]
\centerline{\includegraphics[width=\textwidth]{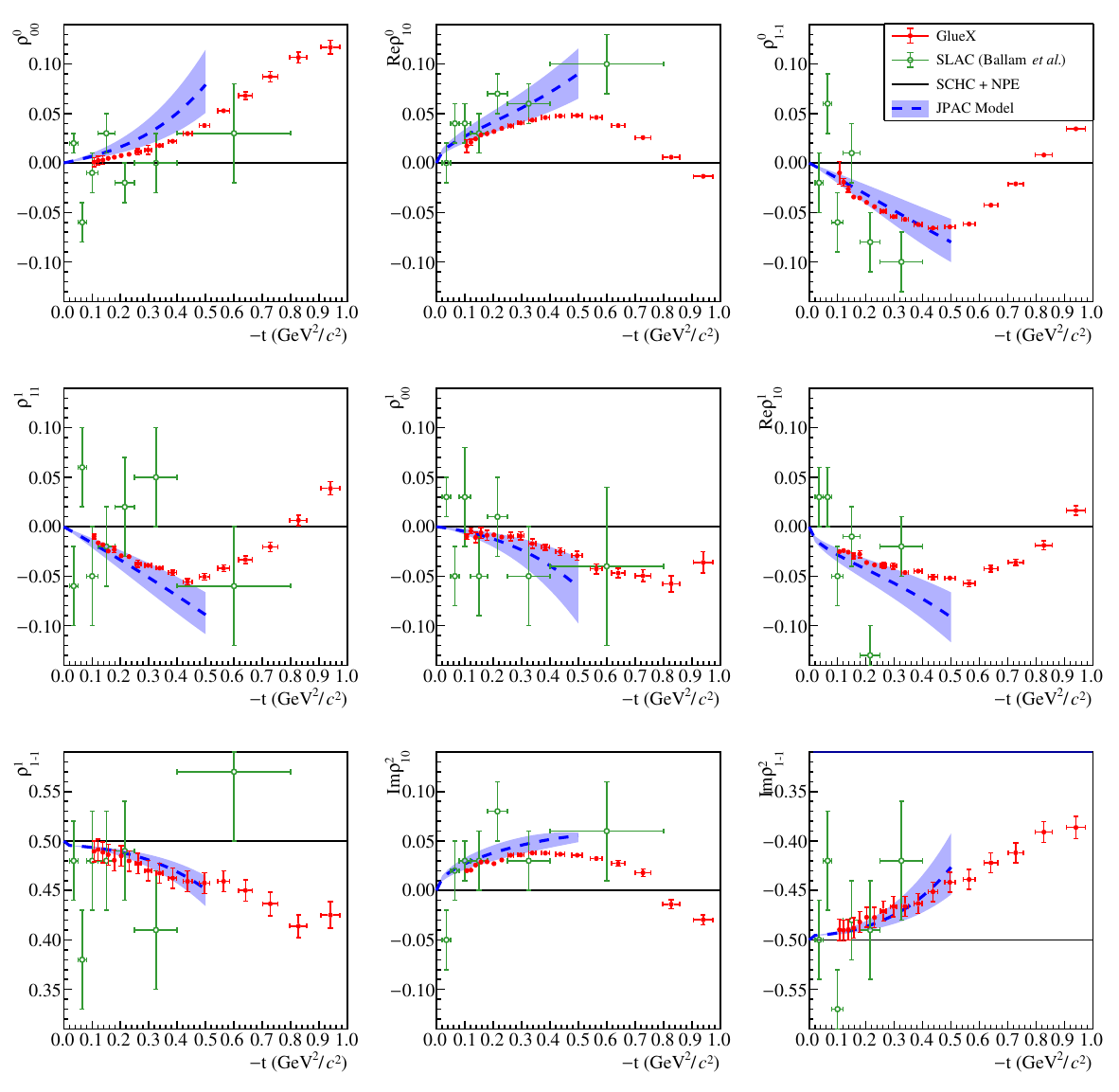}}
\caption{Spin-density matrix elements for the photoproduction of $\rho(770)$ in the helicity system. Our results are shown in red, the error bars display the statistical and systematic uncertainties added in quadrature. The systematic uncertainties for the polarized SDMEs $\rho^1_{ij}$ and $\rho^2_{ij}$ contain an overall relative polarization uncertainty of $2.1\%$ which is fully correlated for all values of $t$. The earlier results from SLAC\cite{Ballam:1972eq} are shown in green. The horizontal black lines show the values for $s$-channel helicity conservation with natural parity exchange (SCHC + NPE), while the blue dashed curves show Regge theory predictions from JPAC with shaded, one-standard-deviation uncertainty bands~\cite{Mathieu:2018xyc}. \label{fig:rhosdme}}
\end{figure*}

The analysis is performed in 18 independent bins in $-t$ between $0.1$ and $1.0\,\text{GeV}^2/c^{2}$. The SDMEs obtained are shown in Fig.~\ref{fig:rhosdme}, together with the earlier results from SLAC~\cite{Ballam:1972eq}, the predictions from $s$-channel helicity conservation with natural parity exchange, and from the JPAC model~\cite{Mathieu:2018xyc}. We report the measured SDMEs at the mean value for each $t$ bin and display the standard deviation of the distribution in $t$ within the bin by horizontal error bars. The vertical error bars correspond to the statistical and systematic uncertainties added in quadrature. The numerical values for the data shown in  Fig.~\ref{fig:rhosdme} are listed in Appendix~\ref{sec:appnum} and can be found in Ref.~\cite{supplemental}.

In the limit of $-t \rightarrow 0$, our results are consistent with the SCHC + NPE model (see~Appendix~\ref{sec:appschc}). Deviations from this description are predicted by Regge theory~\cite{Mathieu:2018xyc} and originate from the interplay of leading natural parity exchanges ($\mathbb{P}, f_2$ and $a_2$) and unnatural exchanges (e.g. $\pi$) and their different dependencies on the squared four-momentum transfer~$t$. Our measurements follow the prediction qualitatively up to the point where the prediction loses its validity at around $-t\approx0.5\,\text{GeV}^2/c^{2}$. We are able to extract the SDMEs with high precision up to $-t=1\,\text{GeV}^2/c^{2}$, and we will discuss the observed deviation from the SCHC+NPE model and the Regge theory prediction in more detail in the following sections.

\subsection{Parity-Exchange Components}\label{sec:parityexchange}

\begin{figure*}[pht!]
  \centerline{\includegraphics[width=\textwidth]{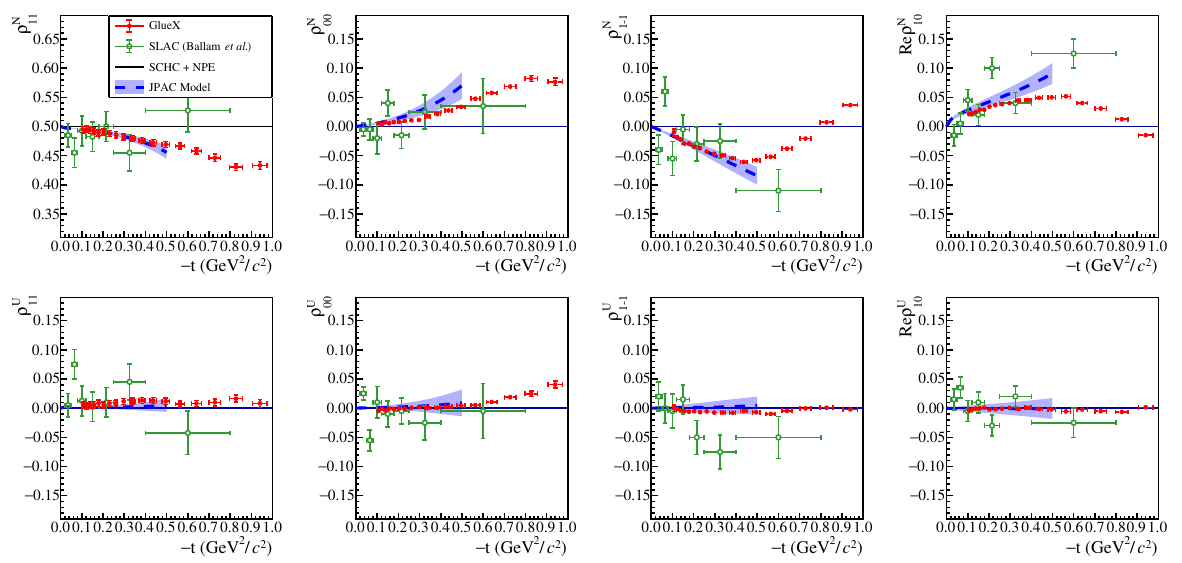}}
  \caption{The spin-density matrix elements for $\rho(770)$ photoproduction for natural- (top row) and unnatural-parity exchange (bottom row). See comments in  Fig.~\ref{fig:rhosdme} caption for details. \label{fig:rhoparity}}
\end{figure*}

The spin-density matrix can be separated into the components $\rho_{ik}^{\mathrm{N,U}}$ arising from natural ($P = (-1)^J$) or unnatural ($P=-(-1)^J$) parity exchanges in the $t$ channel, respectively. The interference term between both production mechanisms vanishes in the limit of high energy~\cite{Schilling:1969um}. We use the results from Fig.~\ref{fig:rhosdme} to calculate the linear combinations
\begin{eqnarray}\label{eq:unnatural}
        \rho^{\mathrm{N,U}}_{ik} = \frac{1}{2} \left [ \rho_{ik}^0 \mp (-1)^i\rho_{-ik}^1 \right ]~.
\end{eqnarray}
Fig.~\ref{fig:rhoparity} illustrates the clean separation. All unnatural components are significantly smaller than their natural counterparts. This means that the deviation from the pure SCHC + NPE model is mainly driven by natural-parity exchange processes, which supports an earlier observation~\cite{Ballam:1972eq}.

\begin{figure}[ht!]
    \centerline{\includegraphics[width=.5\textwidth]{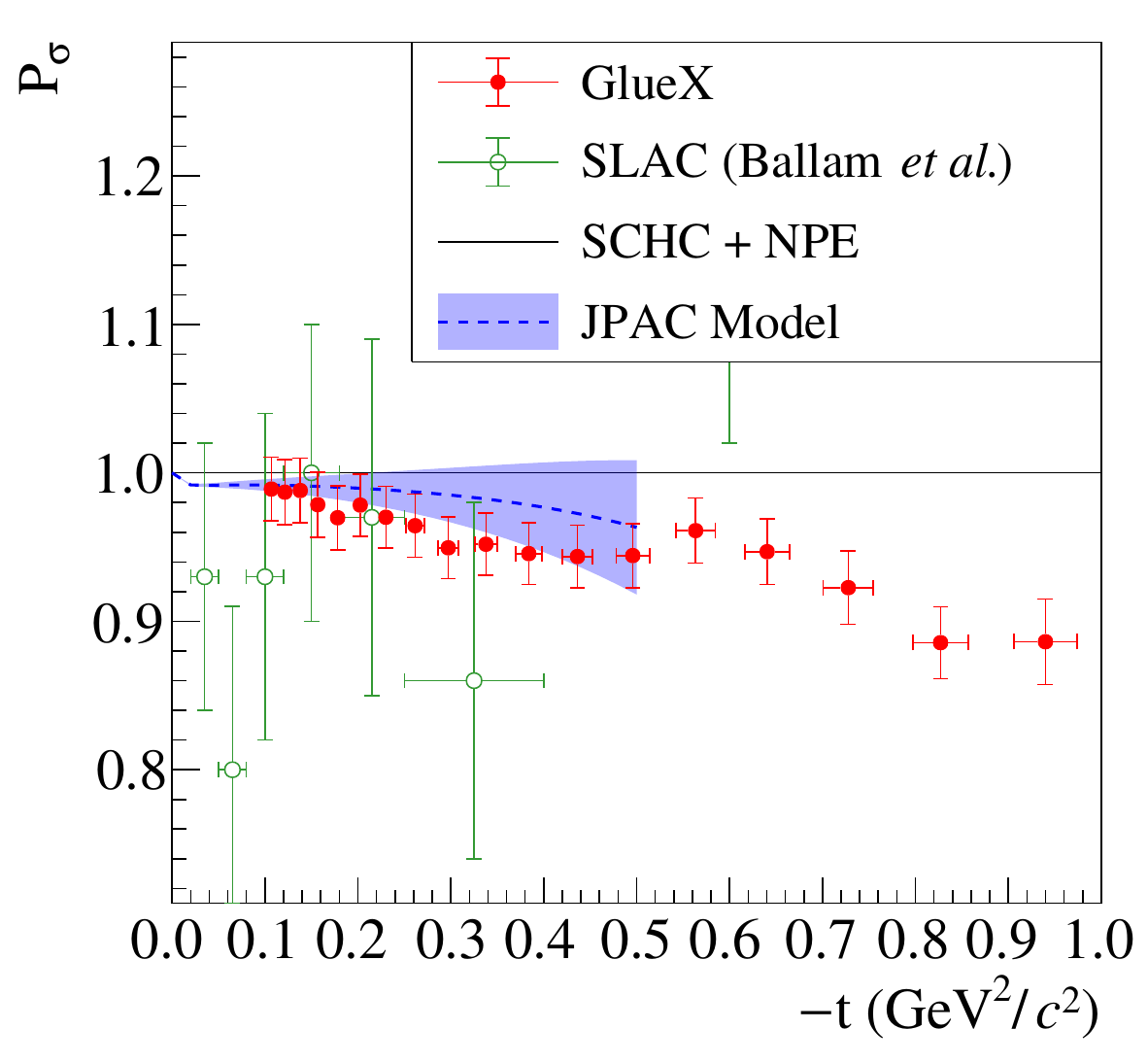}}
    \caption{\label{fig:psigma}Parity asymmetry $P_\sigma$ for $\rho(770)$ photoproduction. See comments in  Fig.~\ref{fig:rhosdme} caption for details.}
\end{figure}

To leading order, the asymmetry between natural- and unnatural-exchange cross sections can be reduced to one single observable, the parity asymmetry $P_\sigma$~\cite{Schilling:1969um}, which is defined as
\begin{eqnarray}\label{eq:psigma}
         P_\sigma = \frac{\sigma^N - \sigma^U}{\sigma^N + \sigma^U} = 2\rho^1_{1-1} - \rho^1_{00}.
\end{eqnarray}
In Fig.~\ref{fig:psigma}, we compare our measured $P_\sigma$ values with previous measurements and the Regge model. For $-t$ below $0.2\,\mathrm{GeV}^2/c^2$, the results are consistent with unity, which again indicates pure natural-parity exchange. The deviation grows towards larger values of $-t$ and is predicted by Regge theory.

\subsection{\label{sec:sdme-relations}Relations between SDMEs}
The spin-density matrix for vector mesons can be written in the center-of-mass frame helicity representation~\cite{Schilling:1969um} as
\begin{eqnarray}
\label{eq:jacob-rel}
\rho(V)_{\lambda_{V}\lambda_{V}'} & = & \frac{1}{\cal N}\,
\sum_{\lambda_{N'}\lambda_{\gamma}\lambda_{N}\lambda_{\gamma}'} \,
T_{\lambda_{V}\lambda_{N'},\lambda_{\gamma}\lambda_{N}}\, \rho(\gamma)_{\lambda_{\gamma}\lambda_{\gamma}'} \,
T^{*}_{\lambda_{V}'\lambda_{N'},\lambda_{\gamma}'\lambda_{N}}\, ,
\end{eqnarray}
where the $\lambda_{x}$ represent the helicity of the incoming ($N$) and outgoing ($N'$) nucleon, the photon ($\gamma$) and the vector meson ($V$), and $T$ is the production amplitude. The term ${\cal N}$ is a normalization factor given as
\begin{eqnarray}
\label{eq:T-norm}
{\cal N} & = & \frac{1}{2}\,  
\sum_{\lambda_{V}\lambda_{N'}\lambda_{\gamma}\lambda_{N}} \, \left |
T_{\lambda_{V}\lambda_{N'},\lambda_{\gamma}\lambda_{N}} \right | ^{2} \, ,
\end{eqnarray}
which for a given center-of-mass momentum $k$ of the incoming photon is related to the unpolarized differential cross section as
\begin{eqnarray}
\frac{d\sigma}{d\Omega} & = & \frac{1}{2}\, \left( \frac{2\pi}{k} \right )^{2}\, {\cal N} \, .
\end{eqnarray}

The $\rho^{\alpha}$ from Eq.~\eqref{eq:rho-alpha} are related to the amplitudes $T$ by
\begin{eqnarray}
    \label{eq:rho0}
\rho^{0}_{\lambda_{V}\lambda_{V}'}(V) & = & \frac{1}{2\cal N}\,
\sum_{\lambda_{N'}\lambda_{\gamma}\lambda_{N}} \,
T_{\lambda_{V}\lambda_{N'},\lambda_{\gamma}\lambda_{N}}\, 
T^{*}_{\lambda_{V}'\lambda_{N'},\lambda_{\gamma}\lambda_{N}}\\
  \label{eq:rho1}
\rho^{1}_{\lambda_{V}\lambda_{V}'}(V) & = & \frac{1}{2\cal N}\,
\sum_{\lambda_{N'}\lambda_{\gamma}\lambda_{N}} \,
T_{\lambda_{V}\lambda_{N'},-\lambda_{\gamma}\lambda_{N}}\, 
T^{*}_{\lambda_{V}'\lambda_{N'},\lambda_{\gamma}\lambda_{N}}\\
    \label{eq:rho2}
\rho^{2}_{\lambda_{V}\lambda_{V}'}(V) & = & \frac{i}{2\cal N}\,
\sum_{\lambda_{N'}\lambda_{\gamma}\lambda_{N}} \,
\lambda_{\gamma}\,T_{\lambda_{V}\lambda_{N'},-\lambda_{\gamma}\lambda_{N}}\, 
T^{*}_{\lambda_{V}'\lambda_{N'},\lambda_{\gamma}\lambda_{N}} \, .
\end{eqnarray}
Thus, the SDMEs are formed from helicity amplitudes that connect the vector-meson helicity $\lambda_{V}$ to the photon helicity $\lambda_{\gamma}$. In the helicity system, $s$-channel helicity conservation implies that the two helicities are equal \mbox{$\lambda_{V}=\lambda_{\gamma}$} (see~Appendix~\ref{sec:appschc}). When SCHC is true, then of the nine measured SDMEs, only $\rho^{1}_{1-1}$ and $\Im\, \rho^{2}_{1-1}$ are nonzero, and $\rho^{1}_{1-1} = -\Im\, \rho^{2}_{1-1}$. If, in addition to SCHC, the production mechanism is described by the exchange of a particle with natural parity in the $t$-channel, then $\rho^{1}_{1-1}=\frac{1}{2}$. If a particle with unnatural parity is exchanged, then $\rho^{1}_{1-1}=-\frac{1}{2}$. As seen in Fig.~\ref{fig:rhosdme}, SCHC + NPE is only valid near $-t=0$.

Going beyond the case where \mbox{ $\lambda_{V}=\lambda_{\gamma}$ }, there could also be amplitudes in which the helicity changes by one or even two units. While the former are very likely to occur, we would expect that the latter are suppressed. If we assume that the amplitudes with \mbox{ $\lambda_{V}=\lambda_{\gamma}\pm 2$ } are zero, additional relations between SDMEs should hold, i.e. Eqs.~\eqref{eq:compare-1m11}, \eqref{eq:compare-101} and~\eqref{eq:compare-rho-0}.
\begin{eqnarray}
\Im\, \rho^{2}_{1-1}  & = & - \rho^{1}_{1-1} \label{eq:compare-1m11}  \\
\Im\, \rho^{2}_{10}  & = & - \Re\, \rho^{1}_{10}  \label{eq:compare-101} \\
\Re\, \rho^{0}_{10}  & = & \pm \Re\, \rho^{1}_{10}  \label{eq:compare-rho-0}
\end{eqnarray}

\begin{figure}[ht!]\centering
\includegraphics[width=.5\textwidth]{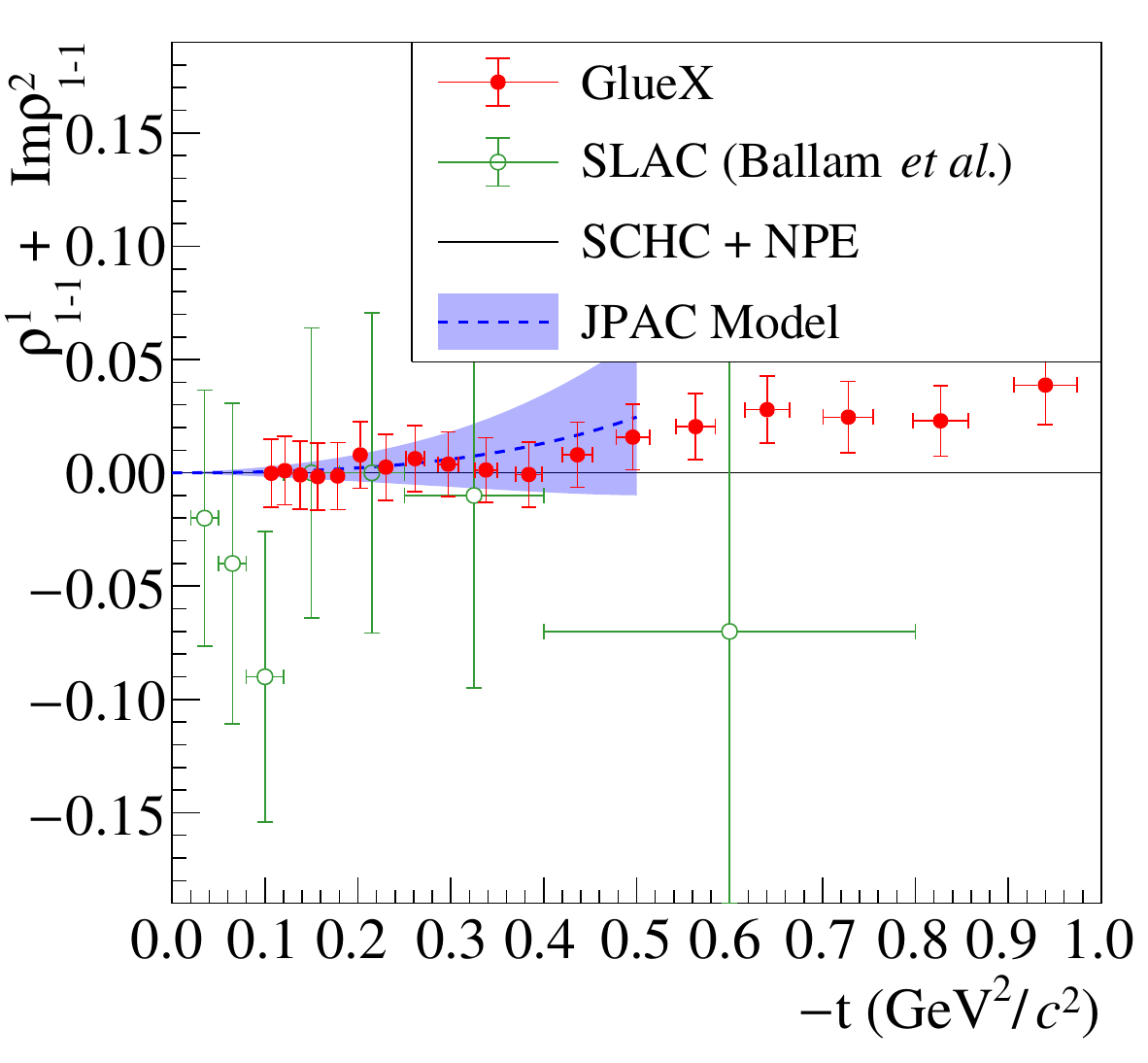}
\caption{The sum of $\rho^1_{1-1}$ and $\Im\, \rho^2_{1-1}$ for $\rho(770)$ photoproduction as a function of $-t$. See comments in  Fig.~\ref{fig:rhosdme} caption for details.}
\label{fig:rho_1m11_1m12}
\end{figure}

To prove Eq.~\eqref{eq:compare-1m11}, we expand Eqs.~\eqref{eq:rho1} and \eqref{eq:rho2} as follows:
\begin{eqnarray}
\rho^{1}_{1-1} & = & \frac{1}{2\cal N}\, \sum_{\lambda_{N}\lambda_{N'}}  \left[ \underbrace{T_{+1\lambda_{N'};+1\lambda_{N}}\,T^{*}_{-1\lambda_{N'};-1\lambda_{N}}}_{\lambda_{\gamma}=-1}\right . \nonumber \\ & & +  \left. \underbrace{T_{+1\lambda_{N'};-1\lambda_{N}}\,T^{*}_{-1\lambda_{N'};+1\lambda_{N}}}_{\lambda_{\gamma}=+1}\right] 
\end{eqnarray}
\begin{eqnarray}
\rho^{2}_{1-1} & = & \frac{i}{2\cal N}\,\sum_{\lambda_{N}\lambda_{N'}} \left[ (-1)\, \underbrace{T_{+1\lambda_{N'};+1\lambda_{N}}\,T^{*}_{-1\lambda_{N'};-1\lambda_{N}}}_{\lambda_{\gamma}=-1} \right . \nonumber \\ & & + \left. (+1)\,\underbrace{T_{+1\lambda_{N'};-1\lambda_{N}}\,T^{*}_{-1\lambda_{N'};+1\lambda_{N}}}_{\lambda_{\gamma}=+1}\right] ~.
\end{eqnarray}
If we define the first sum in both equations as $A$, and the second as $B$, then we have
\begin{eqnarray}
  \rho^{1}_{1-1} & = & A + B  \\
  \rho^{2}_{1-1} & = & -iA + iB \, .
\end{eqnarray}
Looking more closely at the $A$ and $B$ amplitudes, $A$ only includes terms where the photon helicity and the vector-meson helicity are the same, i.e. $\lambda_{\gamma} = \lambda_{V}$, while $B$ only contains terms where the photon helicity and the vector-meson helicity differ by $2$, which we assume to vanish. Taking $B=0$ we have
\begin{eqnarray}
  \rho^{1}_{1-1} & = & A  \\
  \rho^{2}_{1-1} & = & -iA  \, ,
\end{eqnarray}
which yields Eq.~\eqref{eq:compare-1m11}.
Figure~\ref{fig:rho_1m11_1m12} shows $\rho^{1}_{1-1} + \Im\rho^{2}_{1-1}$ as a function of $-t$, for both the \gluex data and the older SLAC data~\cite{Ballam:1972eq}. The sum is consistent with zero for $-t$ values up to about $0.5\, \text{GeV}^2/c^{2}$ and becomes slightly positive above that. The JPAC model~\cite{Mathieu:2018xyc} agrees with this prediction over its range of validity. This suggests that amplitudes with $\lambda_{V}=\lambda_{\gamma}\pm2$ may start to become relevant for values of $-t$ larger than $0.5\,\text{GeV}^2/c^{2}$.

\begin{figure}[ht!]\centering
\includegraphics[width=.5\textwidth]{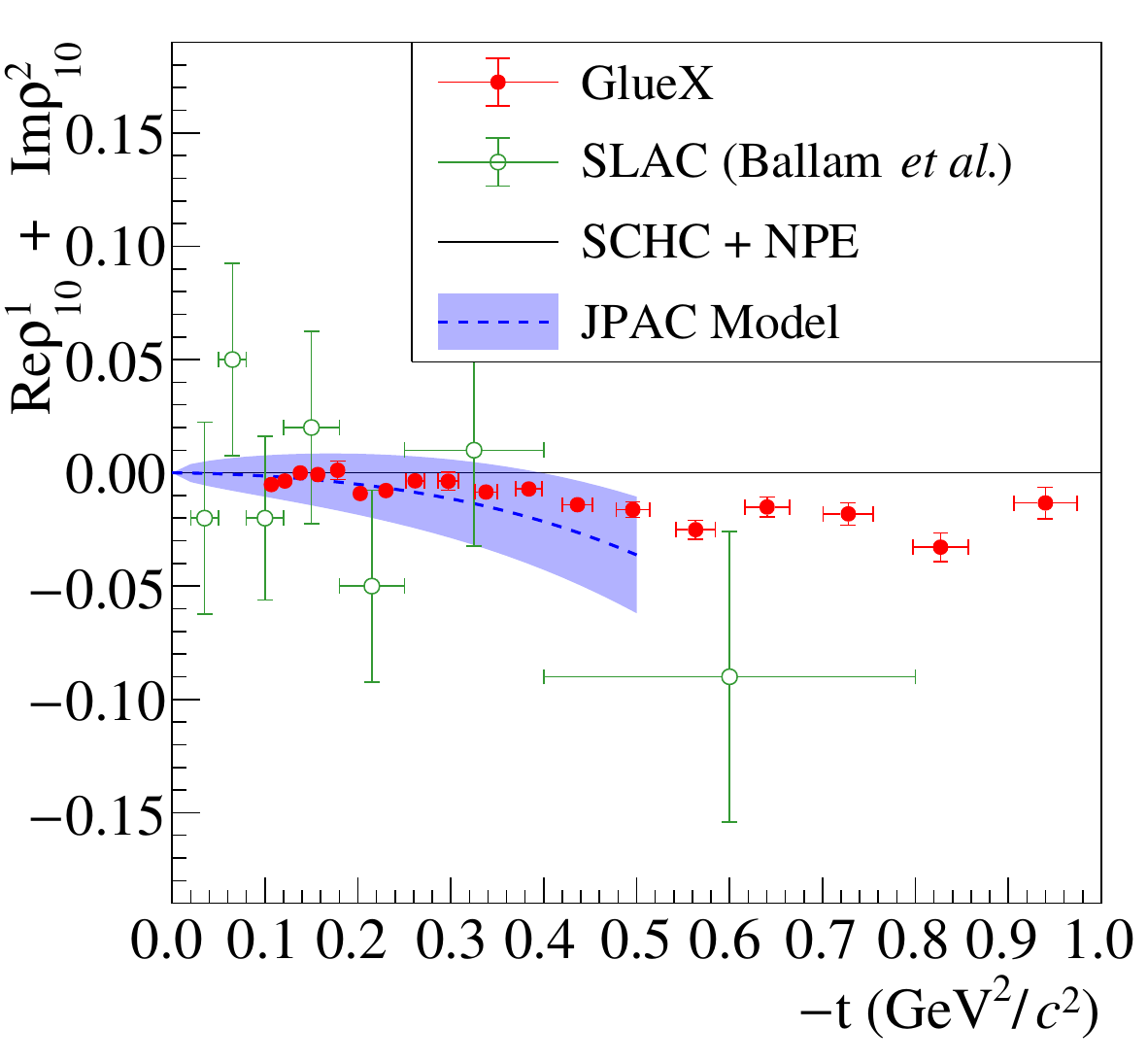}
\caption{The sum of $\Re\rho^0_{10}$ and $\Im\, \rho^2_{10}$ for $\rho(770)$ photoproduction as a function of $-t$. See comments in  Fig.~\ref{fig:rhosdme} caption for details.}
\label{fig:rho_101_102}
\end{figure}

To derive Eq.~\eqref{eq:compare-101}, we perform a similar expansion to the one above:
\begin{eqnarray}
\rho^{1}_{10} & = & \frac{1}{2\cal N}\,\sum_{\lambda_{N}\lambda_{N'}} \left[ \underbrace{T_{+1\lambda_{N'};+1\lambda_{N}}\,T^{*}_{0\lambda_{N'};-1\lambda_{N}}}_{\lambda_{\gamma}=-1} \right . \nonumber  \\
\label{eq:rho-1-1-0} & & + \left.  \underbrace{T_{+1\lambda_{N'};-1\lambda_{N}}\,T^{*}_{0\lambda_{N'};+1\lambda_{N}}}_{\lambda_{\gamma}=+1}\right] \\
\rho^{2}_{10} & = & \frac{i}{2\cal N}\,\sum_{\lambda_{N}\lambda_{N'}} \left[ (-1)\, \underbrace{T_{+1\lambda_{N'};+1\lambda_{N}}\,T^{*}_{0\lambda_{N'};-1\lambda_{N}}}_{\lambda_{\gamma}=-1} \right. \nonumber \\ 
\label{eq:rho-2-1-0} & & + \left .(+1)\, \underbrace{T_{+1\lambda_{N'};-1\lambda_{N}}\,T^{*}_{0\lambda_{N'};+1\lambda_{N}}}_{\lambda_{\gamma}=+1}\right] \, .
\end{eqnarray}
If we define the first sum in both equations as $C$, and the second as $D$, then we have
\begin{eqnarray}
  \rho^{1}_{10} & = & C + D  \\
  \rho^{2}_{10} & = & -iC + iD \, .
\end{eqnarray}
$C$ is an interference term between an amplitude where the photon helicity and the vector-meson helicity are the same, i.e. $\lambda_{\gamma} = \lambda_{V}$, and an amplitude where these helicities differ by $1$. Amplitude $D$ is an interference term between an amplitude where the photon helicity and the vector-meson helicity differ by $1$ and an amplitude where they differ by $2$. Setting the amplitudes that have $\Delta\lambda=2$ to zero gives $D=0$, and consequently yields Eq.~\eqref{eq:compare-101}.
Figure~\ref{fig:rho_101_102} shows the sum of $\Re\rho^{1}_{10}$ and $\Im\, \rho^{2}_{10}$ as a function of $-t$, both for the \gluex data and for the older SLAC data~\cite{Ballam:1972eq}. Comparisons are also made to the JPAC model~\cite{Mathieu:2018xyc}. For the \gluex data, the relationship in Eq.~\eqref{eq:compare-101} appears to be valid for $-t$ below $0.3\,\text{GeV}^2/c^{2}$, where the JPAC model also confirms the relationship. For the \gluex data above $-t$ of $0.5\,\text{GeV}^2/c^{2}$, the sum becomes slightly negative and agrees with the previous observation that amplitudes with $\lambda_{V}=\lambda_{\gamma}\pm2$ may be nonzero for larger values of $-t$.

\begin{figure}[ht!]\centering
\includegraphics[width=.5\textwidth]{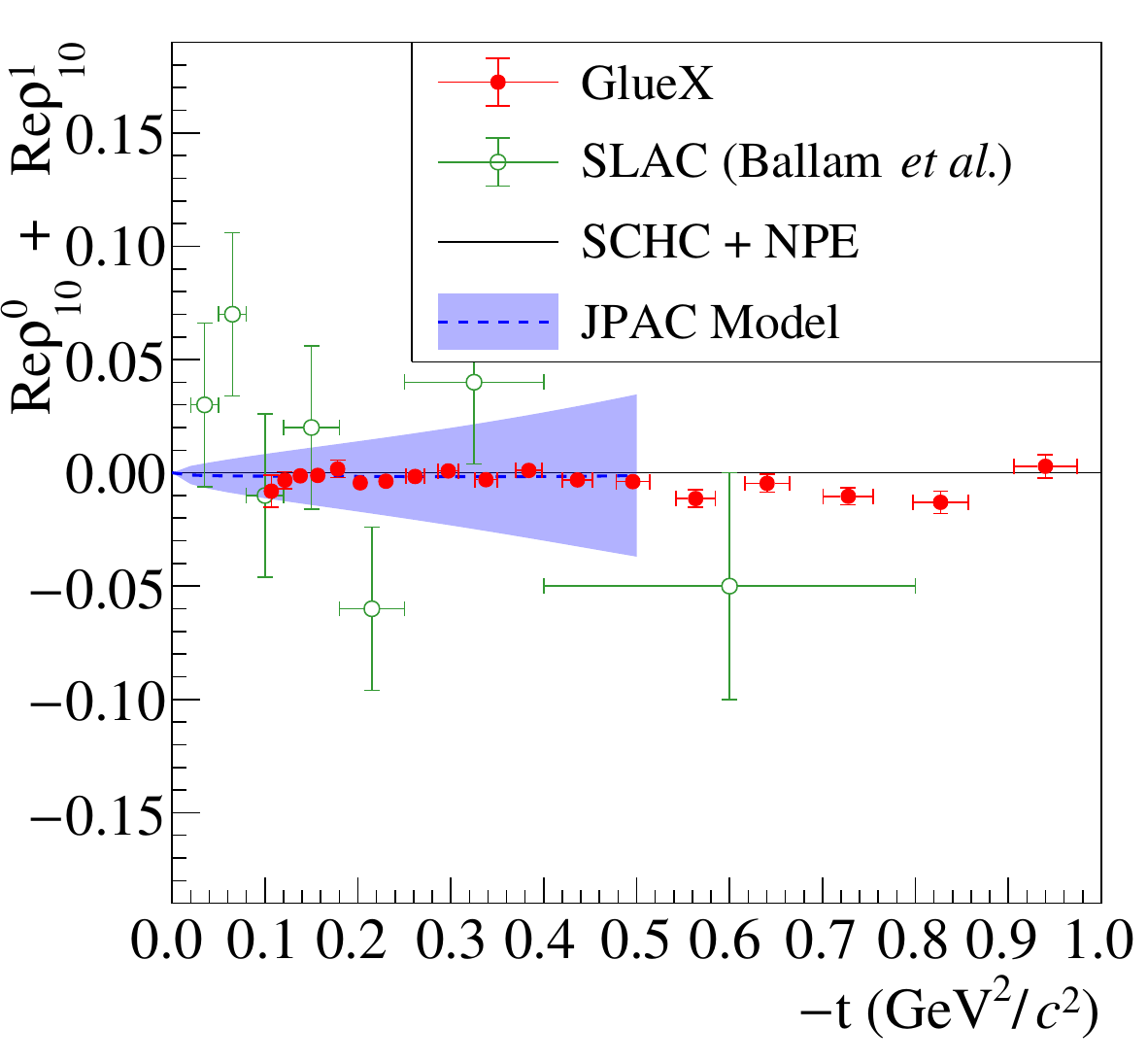}
\caption{The sum of $\Re\rho^0_{10}$ and $\Re\, \rho^1_{10}$ for $\rho(770)$ photoproduction as a function of $-t$. See comments in  Fig.~\ref{fig:rhosdme} caption for details.}
\label{fig:rho_010_101}
\end{figure}

To explain Eq.~\eqref{eq:compare-rho-0}, we write Eq.~\eqref{eq:rho0} as
\begin{eqnarray}
\rho^{0}_{10} & = & \frac{1}{2\cal N}\,\sum_{\lambda_{N}\lambda_{N'}} \left[ \underbrace{T_{+1\lambda_{N'};-1\lambda_{N}}\,T^{*}_{0\lambda_{N'};-1\lambda_{N}}}_{\lambda_{\gamma}=-1} \right . \nonumber \\ & & + \left. \underbrace{T_{+1\lambda_{N'};+1\lambda_{N}}\, T^{*}_{0\lambda_{N'};+1\lambda_{N}}}_{\lambda_{\gamma}=+1}\right] \, .
\label{eq:rho-0-1-0}
\end{eqnarray}
The first term in Eq.~\eqref{eq:rho-0-1-0} describes the interference between an amplitude with $\Delta\lambda=1$ and one with $\Delta\lambda=2$, the latter of which we take to be zero. The second term differs from the first term in Eq.~\eqref{eq:rho-1-1-0} through the difference between the amplitudes $T^{*}_{0\lambda_{N'};-1\lambda_{N}}$ and $T^{*}_{0\lambda_{N'};+1\lambda_{N}}$. These amplitudes connect photons of helicity $\lambda_{\gamma}=\mp1$ to a vector meson of helicity $\lambda_{V}=0$ and only differ by Clebsch-Gordan coefficients that have the same magnitude. For a production mechanism described by a single process, the two amplitudes should be equal in magnitude but could have opposite signs.  Taking $D=0$ and assuming there is a single diagram, then we can write
\begin{eqnarray}
\rho^{0}_{10} & = & \pm C  \, .
\end{eqnarray}
Together with Eq.~\eqref{eq:rho-1-1-0}, this yields Eq.~\eqref{eq:compare-rho-0}.
Figure~\ref{fig:rho_010_101} shows the sum of $\Re\, \rho^{0}_{10}$ and $\Re\, \rho^{1}_{10}$ as a function of $-t$ both for the \gluex data and for the older SLAC data~\cite{Ballam:1972eq}. Comparisons are also made to the JPAC model~\cite{Mathieu:2018xyc}. The \gluex data are consistent with the sum being zero over the full range of $-t$. This suggests that the $\lambda_{V}=\lambda_{\gamma}\pm2$ amplitudes are not important in this case, and that the production mechanism is dominated by a single process, or a series of processes that all contribute with the same sign. The JPAC model also agrees with this prediction.

\section{Conclusions}

We report measurements of the spin-density matrix elements of the $\pi^+\pi^-$ system in the mass range of the vector meson $\rho$(770) ($0.60$ to $0.88\,\text{GeV}/c^2$) photoproduced off the proton with the \gluex experiment at Jefferson Lab. Using a linearly polarized photon beam with energy between $8.2$ and $8.8\,\text{GeV}$ and polarization  close to $35\%$, we reach a statistical precision which surpasses previous measurements by orders of magnitude. The uncertainties on the measurement are dominated by systematic uncertainties, which are studied in detail. Using the full GlueX data set would increase the size of the signal sample five-fold, but would likely not improve the precision of the results further. 

Our results agree well with a prediction by the JPAC collaboration, which was previously fitted to far inferior data. This comparison demonstrates impressively that the description of the production mechanism via a combination of different Regge exchanges is valid at this energy. In particular, the photoproduction of the $\rho(770)$ meson is sensitive to the interplay between Pomeron and $f_2/a_2$ exchanges.

The decomposition of the spin-density matrix elements shows that natural-parity exchanges dominate the production process and that the contribution from unnatural-parity exchanges is small for the analyzed range in squared four-momentum transfer $0.1 < -t < 1.0\,\text{GeV}^2/c^2$. This observation is consistent with the prediction from Regge theory, and the measurements will be used to improve the theoretical description of the reaction.
Based on assumptions about the production process, we predict several relations between the SDMEs and show that these relations are fulfilled by our measurements. In particular, the results strongly suggest that $\rho(770)$ photoproduction at these energies is dominated by a single production mechanism and that contributions from processes where the helicities of the vector meson and the photon differ by two units are negligible.

In this manuscript, we describe the $\pi^+\pi^-$ system with the spin-density matrix elements for a pure $\rho(770)$ meson, but the precision of the data allows us to observe effects that go beyond this simplified picture. The interference with an underlying $S$-wave production of the di-pion system likely influences the SDMEs within the studied mass range. In the future, we plan to study this mass dependence by separating the spin contributions into their individual amplitudes. The formalism outlined in~\cite{Mathieu:2019} will serve as the basis for this investigation.

\section{Acknowledgments}
The analysis in this article was supported by the U.S. Department of Energy, Office of Science, Office of Nuclear Physics under contract DOE Grant No. DE-FG02-87ER40315. The work of A.A. was supported by the DOE, Office of Science, Office of Nuclear Physics in the Early Career Program. We would like to acknowledge the outstanding efforts of the staff of the Accelerator and the Physics Divisions at Jefferson Lab that made the experiment possible. This work was also supported in part by the U.S. Department of Energy, the U.S. National Science Foundation, NSERC Canada, the German Research Foundation, GSI Helmholtzzentrum f\"{u}r Schwerionenforschung GmbH, the Russian Foundation for Basic Research, the UK Science and Technology Facilities Council, the Chilean Comisi\'{o}n Nacional de Investigaci\'{o}n Cient\'{i}fica y Tecnol\'{o}gica, the National Natural Science Foundation of China, and the China Scholarship Council. This material is based upon work supported by the U.S. Department of Energy, Office of Science, Office of Nuclear Physics under contract DE-AC05-06OR23177.
\bibliographystyle{unsrt}
\bibliography{rho_SDME}

\providecommand{\noopsort}[1]{}\providecommand{\singleletter}[1]{#1}%
\begin{thebibliography}{10}

\bibitem{Sakurai:1960ju}
J.~J. Sakurai.
\newblock {Theory of strong interactions}.
\newblock {\em Annals Phys.}, 11:1--48, 1960.

\bibitem{Gilman:1970vi}
Frederick~J. Gilman, Jon Pumplin, A.~Schwimmer, and Leo Stodolsky.
\newblock {Helicity Conservation in Diffraction Scattering}.
\newblock {\em Phys. Lett. B}, 31:387--390, 1970.

\bibitem{Harari:1970fw}
H.~Harari and Y.~Zarmi.
\newblock {Helicity conservation in diffraction scattering and duality}.
\newblock {\em Phys. Lett. B}, 32:291--293, 1970.

\bibitem{Bialas:1971xk}
A.~Bialas, J.~Dabkowski, and L.~Van~Hove.
\newblock {Helicity conservation and the Gribov-Morrison parity rule in
  diffractive pion production}.
\newblock {\em Nucl. Phys. B}, 27:338--348, 1971.

\bibitem{Schopper:1988hwx}
A.~Baldini, V.~Flaminio, W.~G. Moorhead, and Douglas R.~O. Morrison.
\newblock {\em {Total Cross-Sections for Reactions of High Energy Particles
  (Including Elastic, Topological, Inclusive and Exclusive Reactions)}}, volume
  12b of {\em Landolt-Boernstein - Group I Elementary Particles, Nuclei and
  Atoms}.
\newblock Springer, 1988.

\bibitem{Schilling:1969um}
K.~Schilling, P.~Seyboth, and Guenter~E. Wolf.
\newblock {On the Analysis of Vector Meson Production by Polarized Photons}.
\newblock {\em Nucl.\ Phys.\ B}, 15:397--412, 1970.
\newblock [Erratum: Nucl.Phys.B 18, 332 (1970)].

\bibitem{Pichowsky:1994gh}
Michael Pichowsky, Cetin Savkli, and Frank Tabakin.
\newblock {Polarization observables in vector meson photoproduction}.
\newblock {\em Phys. Rev. C}, 53:593--610, 1996.

\bibitem{Kloet:1998js}
W.~M. Kloet, Wen-Tai Chiang, and Frank Tabakin.
\newblock {Spin information from vector - meson decay in photoproduction}.
\newblock {\em Phys. Rev. C}, 58:1086--1097, 1998.

\bibitem{Kloet:1999wc}
W.~M. Kloet and Frank Tabakin.
\newblock {Constraints on vector meson photoproduction spin observables}.
\newblock {\em Phys. Rev. C}, 61:015501, 1999.

\bibitem{Criegee:1968fxl}
L.~Criegee et~al.
\newblock {$\rho$ production with polarized photons}.
\newblock {\em Phys. Lett. B}, 28:282--286, 1968.

\bibitem{Diambrini-Palazzi:1970jgv}
G.~Diambrini-Palazzi, G.~Mcclellan, Nari~B. Mistry, P.~Mostek, H.~Ogren,
  J.~Swartz, and R.~Talman.
\newblock {Photoproduction of rho mesons from hydrogen and carbon by linearly
  polarized photons}.
\newblock {\em Phys. Rev. Lett.}, 25:478--481, 1970.

\bibitem{Ballam:1970qn}
Joseph Ballam et~al.
\newblock {Conservation of $s$ channel helicity in $\rho^0$ photoproduction}.
\newblock {\em Phys. Rev. Lett.}, 24:960--963, 1970.

\bibitem{Ballam:1971yd}
Joseph Ballam et~al.
\newblock {Bubble Chamber Study of Photoproduction by 2.8-GeV and 4.7-GeV
  Polarized Photons. 1. Cross-Section Determinations and Production of $\rho^0$
  and $\Delta^{++}$ in the Reaction $\gamma p \rightarrow p \pi^+ \pi^-$}.
\newblock {\em Phys. Rev. D}, 5:545, 1972.

\bibitem{Ballam:1972eq}
Joseph Ballam et~al.
\newblock {Vector Meson Production by Polarized Photons at 2.8-GeV, 4.7-GeV,
  and 9.3-GeV}.
\newblock {\em Phys.\ Rev.\ D}, 7:3150, 1973.

\bibitem{Bonn-CERN-EcolePoly-Glasgow-Lancaster-Manchester-Orsay-Paris-Rutherford-Sheffield:1982qiv}
D.~Aston et~al.
\newblock {Photoproduction of $\rho^0$ and $\omega$ on Hydrogen at Photon
  Energies of 20-{GeV} to 70-{GeV}}.
\newblock {\em Nucl. Phys. B}, 209:56--76, 1982.

\bibitem{SLACHybridFacilityPhoton:1984hkg}
K.~Abe et~al.
\newblock {Test of S Channel Helicity Conservation in Inelastic $\rho^0$
  Diffraction in 20-{GeV} Photoproduction}.
\newblock {\em Phys. Rev. D}, 32:2288, 1985.

\bibitem{Mathieu:2018xyc}
V.~Mathieu, J.~Nys, C.~Fern\'andez-Ram\'\i{}rez, A.~Jackura, A.~Pilloni,
  N.~Sherrill, A.~P. Szczepaniak, and G.~Fox.
\newblock {Vector Meson Photoproduction with a Linearly Polarized Beam}.
\newblock {\em Phys. Rev. D}, 97(9):094003, 2018.

\bibitem{Austregesilo:2019tld}
Alexander Austregesilo.
\newblock {Spin-density matrix elements for vector meson photoproduction at
  GlueX}.
\newblock {\em AIP Conf. Proc.}, 2249(1):030005, 2020.

\bibitem{GlueX:2020idb}
S.~Adhikari et~al.
\newblock {The GLUEX beamline and detector}.
\newblock {\em Nucl. Instrum. Meth. A}, 987:164807, 2021.

\bibitem{Pooser:2019rhu}
E.~Pooser et~al.
\newblock {The GlueX Start Counter Detector}.
\newblock {\em Nucl. Instrum. Meth. A}, 927:330--342, 2019.

\bibitem{Jarvis:2019mgr}
N.~S. Jarvis et~al.
\newblock {The Central Drift Chamber for GlueX}.
\newblock {\em Nucl. Instrum. Meth. A}, 962:163727, 2020.

\bibitem{Pentchev2017281}
L.~Pentchev et~al.
\newblock {Studies with cathode drift chambers for the GlueX experiment at
  Jefferson Lab}.
\newblock {\em Nucl. Instrum. Meth.}, 845:281 -- 284, 2017.
\newblock Proceedings of the Vienna Conference on Instrumentation 2016.

\bibitem{Beattie:2018xsk}
T.~D. Beattie et~al.
\newblock {Construction and Performance of the Barrel Electromagnetic
  Calorimeter for the GlueX Experiment}.
\newblock {\em Nucl. Instrum. Meth. A}, 896:24--42, 2018.

\bibitem{Dugger:2017zoq}
M.~Dugger et~al.
\newblock {Design and construction of a high-energy photon polarimeter}.
\newblock {\em Nucl. Instrum. Meth. A}, 867:115--127, 2017.

\bibitem{GlueX:2017zoo}
H.~Al~Ghoul et~al.
\newblock {Measurement of the beam asymmetry $\Sigma$ for $\pi^0$ and $\eta$
  photoproduction on the proton at $E_\gamma = 9$ GeV}.
\newblock {\em Phys. Rev. C}, 95(4):042201, 2017.

\bibitem{GlueX:2019adl}
S.~Adhikari et~al.
\newblock {Beam Asymmetry $\Sigma$ for the Photoproduction of $\eta$ and
  $\eta^{\prime}$ Mesons at $E_{\gamma}=8.8$ GeV}.
\newblock {\em Phys. Rev. C}, 100(5):052201, 2019.

\bibitem{GlueX:2020qat}
S.~Adhikari et~al.
\newblock {Measurement of the photon beam asymmetry in $\vec{\gamma} p\to
  K^+\Sigma^0$ at $E_{\gamma} = 8.5$ GeV}.
\newblock {\em Phys. Rev. C}, 101(6):065206, 2020.

\bibitem{GlueX:2020fam}
S.~Adhikari et~al.
\newblock {Measurement of beam asymmetry for $\pi^-\Delta^{++}$ photoproduction
  on the proton at $E_\gamma$=8.5 GeV}.
\newblock {\em Phys. Rev. C}, 103(2):L022201, 2021.

\bibitem{GlueX:2021pcl}
S.~Adhikari et~al.
\newblock {Measurement of spin density matrix elements in
  \ensuremath{\Lambda}(1520) photoproduction at 8.2\textendash{}8.8 GeV}.
\newblock {\em Phys. Rev. C}, 105(3):035201, 2022.

\bibitem{ParticleDataGroup:2022pth}
R.~L. Workman et~al.
\newblock {Review of Particle Physics}.
\newblock {\em PTEP}, 2022:083C01, 2022.

\bibitem{Soding:1965nh}
P.~Soding.
\newblock {On the Apparent shift of the rho meson mass in photoproduction}.
\newblock {\em Phys. Lett.}, 19:702--704, 1966.

\bibitem{ALLISON2016186}
J.~Allison et~al.
\newblock Recent developments in geant4.
\newblock {\em Nuclear Instruments and Methods in Physics Research Section A:
  Accelerators, Spectrometers, Detectors and Associated Equipment},
  835:186--225, 2016.

\bibitem{Breit:1936zzb}
G.~Breit and E.~Wigner.
\newblock {Capture of Slow Neutrons}.
\newblock {\em Phys. Rev.}, 49:519--531, 1936.

\bibitem{hip72}
F.~Von~Hippel and C.~Quigg.
\newblock {Centrifugal-barrier effects in resonance partial decay widths,
  shapes, and production amplitudes}.
\newblock {\em Phys. Rev. D}, 5:624--638, 1972.

\bibitem{amptools}
{M.R. Shepherd, J. Stevens, R. Mitchell, M. Albrecht and A. Austregesilo}.
\newblock {AmpTools Version 0.14.5}.
\newblock \url{https://doi.org/10.5281/zenodo.5039377}, 2022.

\bibitem{efron1994introduction}
Bradley Efron and Robert~J Tibshirani.
\newblock {\em An introduction to the bootstrap}.
\newblock CRC press, 1994.

\bibitem{supplemental}
HEPData.
\newblock \url{https://www.hepdata.net/record/ins2660186}, 2023.

\bibitem{Mathieu:2019}
V.~Mathieu et~al.
\newblock Moments of angular distribution and beam asymmetries in $\eta\pi^0$
  photoproduction at {GlueX}.
\newblock {\em Phys. Rev. D}, 100(5), 2019.

\bibitem{Jacob:1959at}
M.~Jacob and G.C. Wick.
\newblock {On the General Theory of Collisions for Particles with Spin}.
\newblock {\em Annals Phys.}, 7:404--428, 1959.

\end{thebibliography}
\newpage
\appendix
\section{Discussion of s-Channel Helicity Conservation}
\label{sec:appschc}
In the photoproduction of vector mesons such as the $\rho(770)$, $\omega(782)$ and $\phi(1020)$, the spin of the produced meson is related to the spin of the initial photon through a helicity amplitude $T$. The spin states are typically represented as density matrices $\rho(V)$ and $\rho(\gamma)$ where the relation between the two (following Schilling \emph{et al.}~\cite{Schilling:1969um}) is given by Eq.~\eqref{eq:rhoTgam}. This relation can be expressed in the center-of-mass frame helicity representation~\cite{Jacob:1959at} as in Eq.~\eqref{eq:jacob-rel}, which we repeat here:
\begin{eqnarray}
\label{eq:A-jacob-rel}
\rho_{\lambda_{V}\lambda_{V}'}(V) & = & \frac{1}{\cal N}
\sum_{\lambda_{N'}\lambda_{\gamma}\lambda_{N}\lambda_{\gamma}'} 
T_{\lambda_{V}\lambda_{N'},\lambda_{\gamma}\lambda_{N}}\, \rho_{\lambda_{\gamma}\lambda_{\gamma}'} 
T^{*}_{\lambda_{V}'\lambda_{N'},\lambda_{\gamma}'\lambda_{N}}\, .
\end{eqnarray}
This expression relates an initial photon with helicity $\lambda_{\gamma} =\pm 1$ to a final-state vector meson with helicity $\lambda_{V}=0$ or $\lambda_{V}=\pm 1$. The normalization factor ${\cal N}$ is given by Eq.~\eqref{eq:T-norm}, and $N$ and $N'$ represent the initial and final-state nucleons. For the SDMEs of interest, we can write Eq.~\eqref{eq:A-jacob-rel} as a sum over the photon and initial and final-state nucleons as in Eqs.~\eqref{eq:rho0},~\eqref{eq:rho1} and~\eqref{eq:rho2} (which are reproduced here as Eqs.~\eqref{eq:rho0a},~\eqref{eq:rho1a} and~\eqref{eq:rho2a} for convenience) where the $T_{\lambda_{V}\lambda_{N'},\lambda_{\gamma}\lambda_{N}}$ are the helicity amplitudes. The $\rho^{0}_{ij}$ elements are related to unpolarized photons while $\rho^{1}_{ij}$ and $\rho^{2}_{ij}$ are correspond to linear polarization:
\begin{eqnarray}
    \label{eq:rho0a}
\rho^{0}_{\lambda_{V}\lambda_{V}'}(V) & = & \frac{1}{2\cal N}\,
\sum_{\lambda_{N'}\lambda_{\gamma}\lambda_{N}} \,
T_{\lambda_{V}\lambda_{N'},\lambda_{\gamma}\lambda_{N}}\, 
T^{*}_{\lambda_{V}'\lambda_{N'},\lambda_{\gamma}\lambda_{N}}\\
  \label{eq:rho1a}
\rho^{1}_{\lambda_{V}\lambda_{V}'}(V) & = & \frac{1}{2\cal N}\,
\sum_{\lambda_{N'}\lambda_{\gamma}\lambda_{N}} \,
T_{\lambda_{V}\lambda_{N'},-\lambda_{\gamma}\lambda_{N}}\, 
T^{*}_{\lambda_{V}'\lambda_{N'},\lambda_{\gamma}\lambda_{N}}\\
    \label{eq:rho2a}
\rho^{2}_{\lambda_{V}\lambda_{V}'}(V) & = & \frac{i}{2\cal N}\,
\sum_{\lambda_{N'}\lambda_{\gamma}\lambda_{N}} \,
\lambda_{\gamma}\,T_{\lambda_{V}\lambda_{N'},-\lambda_{\gamma}\lambda_{N}}\, 
T^{*}_{\lambda_{V}'\lambda_{N'},\lambda_{\gamma}\lambda_{N}}\, .
\end{eqnarray}

For $s$-channel helicity conservation, the only nonzero amplitudes have $\lambda_{V}=\lambda_{\gamma}$. All amplitudes involving a change of helicity, i.e. $\lambda_{V}\neq\lambda_{\gamma}$, are zero. Thus, the only SDMEs which have nonzero~\footnote{The SDME elements $\rho^{0}_{-1-1}$, $\rho^{1}_{-11}$ and $\rho^{2}_{-11}$ are also nonzero, but they are related to $\rho^{0}_{11}$, $\rho^{1}_{1-1}$ and $\rho^{2}_{1-1}$, respectively, via parity conservation. For this reason, we do not list them as independent elements.} values are $\rho^{0}_{11}$, $\rho^{1}_{1-1}$ and $\rho^{2}_{1-1}$. Generally, one does not independently report $\rho^{0}_{11}$ as it is related to $\rho^{0}_{00}$ through the fact that the trace of $\rho^{0}$ is $1$, i.e. $2\rho^{0}_{11}+\rho^{0}_{00}=1$ where $\rho^{0}_{00}=0$ under SCHC.
\begin{eqnarray}
\label{eq:rho0-0}
\rho^{0}_{11} & = & \frac{1}{2\cal N}\,\sum_{\lambda_{N}\lambda_{N'}} \left[ \underbrace{T_{+1\lambda_{N'};-1\lambda_{N}}\,T^{*}_{+1\lambda_{N'};-1\lambda_{N}}}_{\lambda_{\gamma}=-1} \right . \\
& &  + \left . \underbrace{T_{+1\lambda_{N'};+1\lambda_{N}}\,T^{*}_{+1\lambda_{N'};+1} \lambda_{N}}_{\lambda_{\gamma}=+1}\right] \\
\rho^{1}_{1-1} & = & \frac{1}{2\cal N}\,\sum_{\lambda_{N}\lambda_{N'}} \left[ \underbrace{T_{+1\lambda '_{N};+1\lambda_{N}}\,T^{*}_{-1\lambda_{N'};-1\lambda_{N}}}_{\lambda_{\gamma}=-1} \right . \\ & & +  \left . \underbrace{T_{+1\lambda_{N'};-1\lambda_{N}}\,T^{*}_{-1\lambda_{N'};+1\lambda_{N}}}_{\lambda_{\gamma}=+1}\right] \\
\rho^{2}_{1-1} & = & \frac{i}{2\cal N}\,\sum_{\lambda_{N}\lambda_{N'}} \left[  \underbrace{ - T_{+1\lambda_{N'};1\lambda_{N}}\,T^{*}_{-1\lambda_{N'};-1\lambda_{N}}}_{\lambda_{\gamma}=-1} \right. \\ & & 
+ \left. \underbrace{T_{+1\lambda_{N'};-1\lambda_{N}}\,T^{*}_{-1\lambda_{N'};+1\lambda_{N}}}_{\lambda_{\gamma}=+1}\right] 
\end{eqnarray}
These equations can be simplified to 
\begin{eqnarray}
\label{eq:rho0-1}
\rho^{0}_{11} & = & \frac{1}{2\cal N}\, \left[ T_{+-}T^{*}_{+-} + T_{++}T^{*}_{++} \right] \\
\rho^{1}_{1-1} & = & \frac{1}{2\cal N}\, \left[ T_{++}T^{*}_{--} + T_{+-}T^{*}_{-+} \right] \\
\rho^{2}_{1-1} & = & \frac{i}{2\cal N}\, \left[ - T_{++}T^{*}_{--} + T_{-+}T^{*}_{-+} \right]
\end{eqnarray}
where the sum over $\lambda_{N}\lambda '_{N}$ is assumed, and where we simplify the notation of the transition amplitudes by putting $\lambda_{V}$ as the first subscript and $\lambda_{\gamma}$ as the second.

Now, noting that only the $T_{++}$ and $T_{--}$ are nonzero, we have 
\begin{eqnarray}
\rho^{0}_{11} & = & \frac{1}{2\cal N}\, T_{++}T^{*}_{++} \\
\rho^{1}_{1-1} & = & \frac{1}{2\cal N}\, T_{++}T^{*}_{--} \\
\rho^{2}_{1-1} & = & \frac{-i}{2\cal N}\, T_{++}T^{*}_{--} \, .
\end{eqnarray}
From this, we immediately see that SCHC implies that $\rho^{1}_{1-1} = -\Im\,\rho^{2}_{1-1}$. We also know that only $\rho^{0}$ has a nonzero trace, i.e. 
\begin{eqnarray}
1 & = & \rho^{0}_{11} + \rho^{0}_{00} + \rho^{0}_{-1-1} \, .   
\end{eqnarray}
However, we have established that $\rho^{0}_{00}=0$ and symmetry gives that $\rho^{0}_{-1-1} = \rho^{0}_{11}$. Thus, we have $\rho^{0}_{11}=\frac{1}{2}$. Similarly, $\rho^{0}_{-1-1}=\frac{1}{2}$. Expanding $\rho^{0}_{-1-1}$ as in Eqs.~\eqref{eq:rho0-0} and~\eqref{eq:rho0-1}, we find
\begin{eqnarray}
    \rho^{0}_{-1-1} & = & \frac{1}{2\cal N}\, T_{--}T^{*}_{--} \, ,
\end{eqnarray}
hence, we have 
\begin{eqnarray}
T_{++}T^{*}_{++} & = & T_{--}T^{*}_{--} \, .
\end{eqnarray}
From this we have that
\begin{eqnarray}
    \frac{1}{2\cal N} T_{++}T^{*}_{++} & = & \frac{1}{2}\, ,
\end{eqnarray}
or the amplitude $T_{++}$ can be expressed in complex polar form as
\begin{eqnarray}
     \label{eq:T++}
    \frac{1}{\sqrt{2\cal N}} T_{++} & = & \frac{1}{\sqrt{2}} e^{i\phi_{+}} \, ,
\end{eqnarray}
where $\phi_{+}$ is some phase associated with the amplitude. Similarly the amplitude $T_{--}$ can be expressed in complex polar form as
\begin{eqnarray}
     \label{eq:T--}
    \frac{1}{\sqrt{2\cal N}} T_{--} & = & \frac{1}{\sqrt{2}} e^{i\phi_{-}} \, ,
\end{eqnarray}
where $\phi_{-}$ is the phase associated with $T_{--}$. Combining Eqs.~\eqref{eq:T++} and~\eqref{eq:T--}, SCHC predicts
\begin{eqnarray}
\rho^{0}_{11} & = & \frac{1}{2} \\
\rho^{1}_{1-1} & = & \frac{1}{2} \cos \left(\phi_{+}-\phi_{-} \right) \\
\Im\, \rho^{2}_{1-1} & = & -\frac{1}{2} \cos \left( \phi_{+}-\phi_{-} \right) \, .
\end{eqnarray}
Thus, the magnitudes and signs of $\rho^{1}_{1-1}$ and $\Im\, \rho^{2}_{1-1}$ depend on the phase difference $\Delta\phi=\phi_{+}-\phi_{-}$. In Section~\ref{sec:parityexchange} we discussed the parity asymmetry $P_{\sigma}$ as given in  Eq.~\eqref{eq:psigma}. For pure natural parity exchange, $P_{\sigma}=1$, while for pure unnatural parity exchange, $P_{\sigma}=-1$. In the case of pure natural parity exchange, we have $\Delta\phi =0$ so $\rho^{1}_{1-1}=\frac{1}{2}$ and $\Im\, \rho^{2}_{1-1} = -\frac{1}{2}$. In the case of pure unnatural parity exchange, $\Delta\phi = \pi$ so $\rho^{1}_{1-1}=-\frac{1}{2}$ and $\Im\, \rho^{2}_{1-1} = \frac{1}{2}$. Throughout this article, we refer to $s$-channel helicity conservation plus natural parity exchange, ``SCHC + NPE'', this assumption implies the case of $\Delta\phi=0$ and implies the following predictions for the nonzero SDMEs
\begin{eqnarray}
\rho^{0}_{11} & = & + \frac{1}{2} \\
\rho^{1}_{1-1} & = & + \frac{1}{2} \\
\Im \, \rho^{2}_{1-1}  & = & - \frac{1}{2} \, .
\end{eqnarray}
\section{Numerical Results}
\label{sec:appnum}

All numerical results for the SDMEs and their statistical and systematic uncertainties are listed in Table~\ref{tab:numbers}.  The systematic uncertainties for the polarized SDMEs $\rho^1_{ij}$ and $\rho^2_{ij}$ contain an overall relative normalization uncertainty of $2.1\%$ which is fully correlated for all values of $t$. Numerical data can also be downloaded from HEPData~\cite{supplemental}.

\begin{table*}[ht]
  %\begin{tiny}
    \centering
    \begin{tabular}{ccccccccccccc}
      \hline\hline
      $-t_{\text{min}}$ &  $-t_{\text{max}}$ & $\overline{-t}$ & $-t_{\text{RMS}}$ & $\rho^0_{00}$ & $\Re\,\rho^0_{10}$ & $\rho^0_{1-1}$ & $\rho^1_{11}$ & $\rho^1_{00}$ & $\Re\,\rho^1_{10}$ & $\rho^1_{1-1}$ & $\Im\,\rho^2_{10}$ & $\Im\,\rho^2_{1-1}$ \\
      \hline
      0.100 & 0.114 & 0.107 & 0.004 & 0.0008 & 0.0171 & -0.0100 & -0.0098 & -0.0101 & -0.0252 & 0.4895 & 0.0200 & -0.4897 \\
      &&&& $\pm$ 0.0003 & $\pm$0.0005 & $\pm$0.0007 & $\pm$0.0020 & $\pm$0.0010 & $\pm$0.0020 & $\pm$0.0024 & $\pm$0.0014 & $\pm$0.0023 \\
      &&&& $\pm$ 0.0045 & $\pm$0.0066 & $\pm$0.0116 & $\pm$0.0016 & $\pm$0.0025 & $\pm$0.0012 & $\pm$0.0103 & $\pm$0.0010 & $\pm$0.0104 \\
      0.114 & 0.129 & 0.121 & 0.004 & 0.0025 & 0.0209 & -0.0194 & -0.0163 & -0.0043 & -0.0242 & 0.4914 & 0.0205 & -0.4904 \\
      &&&& $\pm$ 0.0003 & $\pm$0.0004 & $\pm$0.0006 & $\pm$0.0018 & $\pm$0.0012 & $\pm$0.0017 & $\pm$0.0025 & $\pm$0.0013 & $\pm$0.0022 \\
      &&&& $\pm$ 0.0042 & $\pm$0.0030 & $\pm$0.0038 & $\pm$0.0015 & $\pm$0.0026 & $\pm$0.0014 & $\pm$0.0105 & $\pm$0.0012 & $\pm$0.0103 \\
      0.129 & 0.147 & 0.138 & 0.005 & 0.0030 & 0.0244 & -0.0264 & -0.0182 & -0.0108 & -0.0257 & 0.4886 & 0.0257 & -0.4896 \\
      &&&& $\pm$ 0.0003 & $\pm$0.0004 & $\pm$0.0006 & $\pm$0.0017 & $\pm$0.0010 & $\pm$0.0017 & $\pm$0.0022 & $\pm$0.0011 & $\pm$0.0021 \\
      &&&& $\pm$ 0.0044 & $\pm$0.0023 & $\pm$0.0032 & $\pm$0.0018 & $\pm$0.0052 & $\pm$0.0015 & $\pm$0.0104 & $\pm$0.0011 & $\pm$0.0103 \\
      0.147 & 0.167 & 0.157 & 0.006 & 0.0047 & 0.0283 & -0.0344 & -0.0246 & -0.0061 & -0.0294 & 0.4862 & 0.0287 & -0.4879 \\
      &&&& $\pm$ 0.0002 & $\pm$0.0004 & $\pm$0.0005 & $\pm$0.0017 & $\pm$0.0010 & $\pm$0.0016 & $\pm$0.0023 & $\pm$0.0012 & $\pm$0.0020 \\
      &&&& $\pm$ 0.0022 & $\pm$0.0011 & $\pm$0.0009 & $\pm$0.0018 & $\pm$0.0055 & $\pm$0.0023 & $\pm$0.0103 & $\pm$0.0010 & $\pm$0.0103 \\
      0.167 & 0.190 & 0.178 & 0.007 & 0.0058 & 0.0295 & -0.0353 & -0.0232 & -0.0087 & -0.0278 & 0.4805 & 0.0290 & -0.4819 \\
      &&&& $\pm$ 0.0003 & $\pm$0.0003 & $\pm$0.0006 & $\pm$0.0016 & $\pm$0.0010 & $\pm$0.0017 & $\pm$0.0020 & $\pm$0.0011 & $\pm$0.0022 \\
      &&&& $\pm$ 0.0025 & $\pm$0.0008 & $\pm$0.0008 & $\pm$0.0026 & $\pm$0.0051 & $\pm$0.0034 & $\pm$0.0103 & $\pm$0.0011 & $\pm$0.0101 \\
      0.190 & 0.215 & 0.203 & 0.007 & 0.0075 & 0.0318 & -0.0398 & -0.0294 & -0.0082 & -0.0362 & 0.4850 & 0.0271 & -0.4771 \\
      &&&& $\pm$ 0.0003 & $\pm$0.0004 & $\pm$0.0005 & $\pm$0.0016 & $\pm$0.0012 & $\pm$0.0013 & $\pm$0.0021 & $\pm$0.0011 & $\pm$0.0021 \\
      &&&& $\pm$ 0.0013 & $\pm$0.0013 & $\pm$0.0007 & $\pm$0.0011 & $\pm$0.0010 & $\pm$0.0013 & $\pm$0.0102 & $\pm$0.0007 & $\pm$0.0101 \\
      0.215 & 0.245 & 0.230 & 0.008 & 0.0088 & 0.0349 & -0.0441 & -0.0302 & -0.0105 & -0.0386 & 0.4798 & 0.0308 & -0.4773 \\
      &&&& $\pm$ 0.0003 & $\pm$0.0003 & $\pm$0.0006 & $\pm$0.0017 & $\pm$0.0011 & $\pm$0.0015 & $\pm$0.0022 & $\pm$0.0011 & $\pm$0.0018 \\
      &&&& $\pm$ 0.0012 & $\pm$0.0015 & $\pm$0.0009 & $\pm$0.0011 & $\pm$0.0012 & $\pm$0.0013 & $\pm$0.0101 & $\pm$0.0008 & $\pm$0.0101 \\
      0.245 & 0.278 & 0.262 & 0.010 & 0.0112 & 0.0375 & -0.0488 & -0.0375 & -0.0100 & -0.0391 & 0.4772 & 0.0356 & -0.4710 \\
      &&&& $\pm$ 0.0003 & $\pm$0.0004 & $\pm$0.0006 & $\pm$0.0017 & $\pm$0.0013 & $\pm$0.0016 & $\pm$0.0025 & $\pm$0.0013 & $\pm$0.0021 \\
      &&&& $\pm$ 0.0032 & $\pm$0.0017 & $\pm$0.0007 & $\pm$0.0032 & $\pm$0.0042 & $\pm$0.0024 & $\pm$0.0101 & $\pm$0.0009 & $\pm$0.0099 \\
      0.278 & 0.316 & 0.297 & 0.011 & 0.0132 & 0.0405 & -0.0543 & -0.0391 & -0.0093 & -0.0396 & 0.4701 & 0.0359 & -0.4663 \\
      &&&& $\pm$ 0.0004 & $\pm$0.0004 & $\pm$0.0006 & $\pm$0.0019 & $\pm$0.0014 & $\pm$0.0015 & $\pm$0.0023 & $\pm$0.0011 & $\pm$0.0022 \\
      &&&& $\pm$ 0.0045 & $\pm$0.0006 & $\pm$0.0005 & $\pm$0.0013 & $\pm$0.0039 & $\pm$0.0027 & $\pm$0.0099 & $\pm$0.0022 & $\pm$0.0099 \\
      0.316 & 0.360 & 0.338 & 0.012 & 0.0176 & 0.0433 & -0.0570 & -0.0419 & -0.0171 & -0.0464 & 0.4674 & 0.0379 & -0.4662 \\
      &&&& $\pm$ 0.0004 & $\pm$0.0004 & $\pm$0.0006 & $\pm$0.0019 & $\pm$0.0015 & $\pm$0.0016 & $\pm$0.0029 & $\pm$0.0013 & $\pm$0.0021 \\
      &&&& $\pm$ 0.0024 & $\pm$0.0010 & $\pm$0.0011 & $\pm$0.0015 & $\pm$0.0043 & $\pm$0.0017 & $\pm$0.0098 & $\pm$0.0015 & $\pm$0.0098 \\
      0.360 & 0.409 & 0.384 & 0.014 & 0.0220 & 0.0459 & -0.0622 & -0.0464 & -0.0208 & -0.0449 & 0.4624 & 0.0378 & -0.4631 \\
      &&&& $\pm$ 0.0004 & $\pm$0.0004 & $\pm$0.0008 & $\pm$0.0022 & $\pm$0.0017 & $\pm$0.0017 & $\pm$0.0031 & $\pm$0.0014 & $\pm$0.0027 \\
      &&&& $\pm$ 0.0014 & $\pm$0.0017 & $\pm$0.0012 & $\pm$0.0014 & $\pm$0.0025 & $\pm$0.0010 & $\pm$0.0097 & $\pm$0.0011 & $\pm$0.0097 \\
      0.409 & 0.464 & 0.436 & 0.016 & 0.0297 & 0.0476 & -0.0658 & -0.0557 & -0.0251 & -0.0507 & 0.4592 & 0.0366 & -0.4513 \\
      &&&& $\pm$ 0.0005 & $\pm$0.0005 & $\pm$0.0008 & $\pm$0.0026 & $\pm$0.0020 & $\pm$0.0020 & $\pm$0.0036 & $\pm$0.0017 & $\pm$0.0024 \\
      &&&& $\pm$ 0.0016 & $\pm$0.0015 & $\pm$0.0011 & $\pm$0.0018 & $\pm$0.0033 & $\pm$0.0016 & $\pm$0.0098 & $\pm$0.0012 & $\pm$0.0095 \\
      0.464 & 0.527 & 0.496 & 0.018 & 0.0379 & 0.0480 & -0.0647 & -0.0507 & -0.0293 & -0.0519 & 0.4575 & 0.0356 & -0.4417 \\
      &&&& $\pm$ 0.0006 & $\pm$0.0005 & $\pm$0.0008 & $\pm$0.0029 & $\pm$0.0029 & $\pm$0.0020 & $\pm$0.0042 & $\pm$0.0019 & $\pm$0.0033 \\
      &&&& $\pm$ 0.0022 & $\pm$0.0019 & $\pm$0.0013 & $\pm$0.0017 & $\pm$0.0037 & $\pm$0.0015 & $\pm$0.0097 & $\pm$0.0014 & $\pm$0.0093 \\
      0.527 & 0.599 & 0.564 & 0.021 & 0.0528 & 0.0460 & -0.0617 & -0.0421 & -0.0426 & -0.0574 & 0.4593 & 0.0323 & -0.4389 \\
      &&&& $\pm$ 0.0007 & $\pm$0.0006 & $\pm$0.0011 & $\pm$0.0031 & $\pm$0.0035 & $\pm$0.0027 & $\pm$0.0043 & $\pm$0.0021 & $\pm$0.0038 \\
      &&&& $\pm$ 0.0020 & $\pm$0.0017 & $\pm$0.0015 & $\pm$0.0014 & $\pm$0.0036 & $\pm$0.0022 & $\pm$0.0098 & $\pm$0.0008 & $\pm$0.0093 \\
      0.599 & 0.681 & 0.640 & 0.024 & 0.0681 & 0.0378 & -0.0427 & -0.0334 & -0.0469 & -0.0424 & 0.4500 & 0.0274 & -0.4221 \\
      &&&& $\pm$ 0.0009 & $\pm$0.0008 & $\pm$0.0013 & $\pm$0.0034 & $\pm$0.0043 & $\pm$0.0032 & $\pm$0.0048 & $\pm$0.0025 & $\pm$0.0043 \\
      &&&& $\pm$ 0.0037 & $\pm$0.0018 & $\pm$0.0006 & $\pm$0.0020 & $\pm$0.0023 & $\pm$0.0013 & $\pm$0.0095 & $\pm$0.0015 & $\pm$0.0092 \\
      0.681 & 0.774 & 0.728 & 0.027 & 0.0873 & 0.0257 & -0.0211 & -0.0203 & -0.0496 & -0.0360 & 0.4365 & 0.0179 & -0.4119 \\
      &&&& $\pm$ 0.0012 & $\pm$0.0009 & $\pm$0.0015 & $\pm$0.0046 & $\pm$0.0048 & $\pm$0.0029 & $\pm$0.0074 & $\pm$0.0034 & $\pm$0.0052 \\
      &&&& $\pm$ 0.0051 & $\pm$0.0014 & $\pm$0.0015 & $\pm$0.0015 & $\pm$0.0037 & $\pm$0.0015 & $\pm$0.0094 & $\pm$0.0015 & $\pm$0.0088 \\
      0.774 & 0.880 & 0.827 & 0.030 & 0.1067 & 0.0059 & 0.0080 & 0.0064 & -0.0577 & -0.0189 & 0.4140 & -0.0139 & -0.3910 \\
      &&&& $\pm$ 0.0017 & $\pm$0.0010 & $\pm$0.0020 & $\pm$0.0048 & $\pm$0.0059 & $\pm$0.0041 & $\pm$0.0069 & $\pm$0.0042 & $\pm$0.0064 \\
      &&&& $\pm$ 0.0052 & $\pm$0.0020 & $\pm$0.0014 & $\pm$0.0025 & $\pm$0.0054 & $\pm$0.0017 & $\pm$0.0091 & $\pm$0.0016 & $\pm$0.0084 \\
      0.880 & 1.000 & 0.940 & 0.034 & 0.1170 & -0.0135 & 0.0345 & 0.0388 & -0.0361 & 0.0164 & 0.4251 & -0.0297 & -0.3863 \\
      &&&& $\pm$ 0.0024 & $\pm$0.0012 & $\pm$0.0019 & $\pm$0.0062 & $\pm$0.0078 & $\pm$0.0045 & $\pm$0.0098 & $\pm$0.0049 & $\pm$0.0078 \\
      &&&& $\pm$ 0.0065 & $\pm$0.0016 & $\pm$0.0007 & $\pm$0.0026 & $\pm$0.0074 & $\pm$0.0017 & $\pm$0.0091 & $\pm$0.0012 & $\pm$0.0082 \\
      \hline \hline
    \end{tabular}
    \caption{Spin-density matrix elements for the photoproduction of $\rho(770)$ in the helicity system. The first uncertainty is statistical, the second systematic. The systematic uncertainties for the polarized SDMEs $\rho^1_{ij}$ and $\rho^2_{ij}$ contain an overall relative normalization uncertainty of $2.1\%$ which is fully correlated for all values of $t$.}
    \label{tab:numbers}
    %\end{tiny}
\end{table*}
\end{document}